\documentclass[twocolumn]{svjour3} 

\usepackage[american]{babel}
\usepackage[utf8]{inputenc}
\usepackage{ae,aecompl}
\usepackage[numbers]{natbib}

\usepackage{bm}
\usepackage{url}
\usepackage{color}
\usepackage{booktabs}
\usepackage{graphicx}
\usepackage{subfigure}
\usepackage{float}

\usepackage{amsmath} 
\usepackage{amssymb}
\usepackage{commath}
\usepackage{dsfont}

\newcommand{\R}{\ensuremath{\mathbb{R}}}           

\renewcommand{\vec}[1]{\bm{\mbox{#1}}}

\newcommand{\indfunc}[1]{\ensuremath{\mathds{1}_{#1} }}

\newcommand{\cov}[2]{\texttt{cov} \left(#1, #2\right)}

\newcommand{\varvar}[1]{\texttt{var} \left(#1\right)}

\newcommand{\median}[1]{\texttt{median} \left(#1\right)}

\newcommand{\PM}{\mathcal{P}}

\newcommand{\pdf}[1]{g_{\tiny{#1}}}

\newcommand{\expval}[1]{\mathbb{E}\left\lbrace #1 \right\rbrace}

\newcommand{\estim}[1]{\widehat{#1}}

\journalname{Nonlinear Dynamics}

\begin{document}

\title{Enhancing the performance of a bistable energy harvesting device via the cross-entropy method}

\author{Americo Cunha~Jr}

\institute{A. Cunha~Jr \at
              Rio de Janeiro State University -- UERJ,
              Institute of Mathematics and Statistics,
		     Rio de Janeiro, Brazil\\
		     ORCID: 0000-0002-8342-0363\\
              \email{americo.cunha@uerj.br}
}

\date{Received: date / Accepted: date}

\maketitle

\begin{abstract}
This work deals with the solution of a non-convex optimization problem to enhance the performance of an energy harvesting device, which involves a nonlinear objective function and a discontinuous constraint. This optimization problem, which seeks to find a suitable configuration of parameters that maximize the electrical power recovered by a bistable energy harvesting system, is formulated in terms of the dynamical system response and a binary classifier obtained from 0-1 test for chaos. A stochastic solution strategy that combines penalization and the cross-entropy method is proposed and numerically tested. Computational experiments are conducted to address the performance of the proposed optimization approach by comparison with a reference solution, obtained via an exhaustive search in a refined numerical mesh. The obtained results illustrate the effectiveness and robustness of the cross-entropy optimization strategy (even in the presence of noise or in moderately higher dimensions), showing that the proposed framework may be a very useful and powerful tool to solve optimization problems involving nonlinear energy harvesting dynamical systems.

\end{abstract}

\section{Introduction}

Energy harvesting is a process of energy conversion in which a certain amount of energy is pumped from an abundant source (e.g. the sun, wind, ocean, environmental vibrations, electromagnetic radiation, etc.) into another system that stores and/or use this energy for its self-operation \cite{Gammaitoni2012p627,priya2009,spies2015}. This type of technology has a broad field of applicability, that includes: (i) large-scale devices such as ocean thermal energy converters \cite{Liu2019p109581}, magnetic levitators \cite{Rocha2019p3423}, etc.; (ii) middle size applications like electromechanical systems \cite{Ghouli2017p1625,Leadenham2020p625}, sensors/actuators \cite{Bhatti2016p24,Rade2012p1650,Trindade2016}, living trees \cite{ying2015v7} etc.; (iii) small or very small systems, namely medical implants \cite{pfenniger2014p3325}, micro/nano electro-mechanical systems (MEMS/NENS) \cite{Nabavi2016,Selvan2016p1035}, graphene structures \cite{Gammaitoni2011p161401}, bio cells \cite{Catacuzzeno2019p823}, etc.

Enhance the amount of energy collected by an energy harvester is a key problem in the operation of these kind of devices, being the object of interest of several research works among the last decade, with efforts divided essentially in two fronts: (i) approaches with great physical appeal, which seek to explore complex geometric arrangements and nonlinearities in a smart way to improve the performance of the system\cite{Abdelkefi2012p531,Benacchio2016p893,Daqaq2020p1525,Dekemele2019p1831,harne2012p162,Ibrahim2016,Ramlan2010p545,Vocca2012p771,Zhou2016p1599};
(ii) works that look at this task from a more mathematical perspective, conducting theoretical analysis in mathematical models \cite{rechenbach2016p1232}, using control theory to optimize the underlying dynamics \cite{Savi2015p2787,cunhajr_cobem2019_1,Sun2020,Yang2019p1511} or formulating and solving non-convex optimization problems to find the optimal system design \cite{trindade2012p552,lopes_cnmac2017_2,cunhajr_belhaq2019,peterson_cnmac2016,Zheng2009p17}.

The approaches that seek to optimize energy harvesters exploring their physics, in general, try to take advantage of a key (and very particular) characteristic of the dynamics, often neglected in a naive look, in favor of better performance. Although they can produce very efficient solutions for certain types of systems, they lack generality, which makes them, in a sense, an ``art'', as it requires a high level of knowledge about the system behavior by the designer's part. 

On the other hand, approaches based on numerical optimization have a good level of generality, however, they come up with theoretical and computational difficulties inherent to the solution of non-convex optimization problems defined by objective functions that are constructed from observations of the nonlinear dynamical systems. 

These difficulties are faced not only when dealing with energy harvesters, but they can also be seen in the optimization of several types of nonlinear dynamical systems, generally being circumvented with the use of computational intelligence-based algorithms \cite{Mangla2020,Quaranta2020}, since traditional gradient-based methods have no guarantee of finding global extremes in the absence of convexity \cite{bonnans2009,boyd2004,nocedal2006}.

Among the computational intelligence algorithms available in the literature, the most frequently used for optimization in nonlinear dynamics include genetic algorithms (and their variants), particle swarm optimization, differential evolution, artificial neural networks (and other machine learning methods), etc \cite{Quaranta2020}. As they are global search methods, they are often successful in overcoming the (local) difficulties faced by gradient-based methods, at the price of losing computational efficiency. But the loss of efficiency is not the only weakness of these global methods, most of then need to have the underlying control parameters tuned for proper functioning. This task is usually done manually, in a trial and error fashion, which is not at all practical, as the meaning of many of these parameters is often not intuitive. This peculiarity practically ``condemns'' these tools to be used in the black-box format, without much control by the user, which may induce significant losses in performance and accuracy.

However, the computational intelligence literature has at least one global search algorithm that is robust and simple, for which the control parameters are very intuitive, it is known as the \emph{cross-entropy (CE) method}, proposed by R. Rubinstein in 1997 \cite{Rubinstein1997p89,Rubinstein1999p127,Rubinstein2004} for rare events simulation. Soon after it was realized that it could be a very appealing global search technique for challenging combinatorial and continuous optimization problems \cite{Kroese2013,Rubinstein2016}. It is a sampling technique, from the family of Monte Carlo methods, which iteratively attacks the problem, refining the candidates to be the solution according to a certain optimality criterion.

Surprisingly, the nonlinear dynamics literature has been neglecting the CE method, although it is a relatively known technique in the combinatorial optimization community. To the best of the author's knowledge, there are few studies in the open literature applying the CE method for numerical optimization problems involving nonlinear mechanical systems \cite{Dantas2019,cunhajr_icedyn2019,cunhajr_cobem2019_2,Ghidey2015,cunhajr_cnmac2018,Wang2012}, which leaves space for new contributions in this line. In particular, the development of a simplistic and robust optimization framework, easily customizable, which can be used by naive users without major performance losses.

In this context, this work deals with the numerical optimization of nonlinear energy harvesting systems, trying to find a strategy to increase the performance of a bistable energy harvesting device subjected to a periodic excitation. For this purpose, a  non-convex optimization problem, with a nonlinear objective function and discontinuous constraint, is formulated and a stochastic strategy of solution that combines penalization and the CE method is proposed and numerically tested. As this method has a theoretical guarantee of convergence and evidence in the literature proving its effectiveness \cite{Kroese2013,Rubinstein2016}, the proposed cross-entropy framework for optimization of dynamical systems has the potential to be a very general and robust numerical methodology.

The rest of this manuscript is organized as follows. Section~\ref{non_dynsys} presents the energy harvesting device of interest and the underlying dynamical system. An optimization problem associated with this nonlinear system is defined in section~\ref{nonlinear_opt_prob}, and the CE method strategy of solution is presented in section~\ref{ce_method}. Numerical experiments, conducted to test the effectiveness and robustness of the stochastic solution strategy, are shown in section~\ref{num_experim}. Finally, in section~\ref{concl_sect}, final remarks are set out.

\section{Nonlinear dynamical system}
\label{non_dynsys}

\subsection{Physical system}

\begin{figure}[h!]
\centering
\includegraphics[scale=0.38]{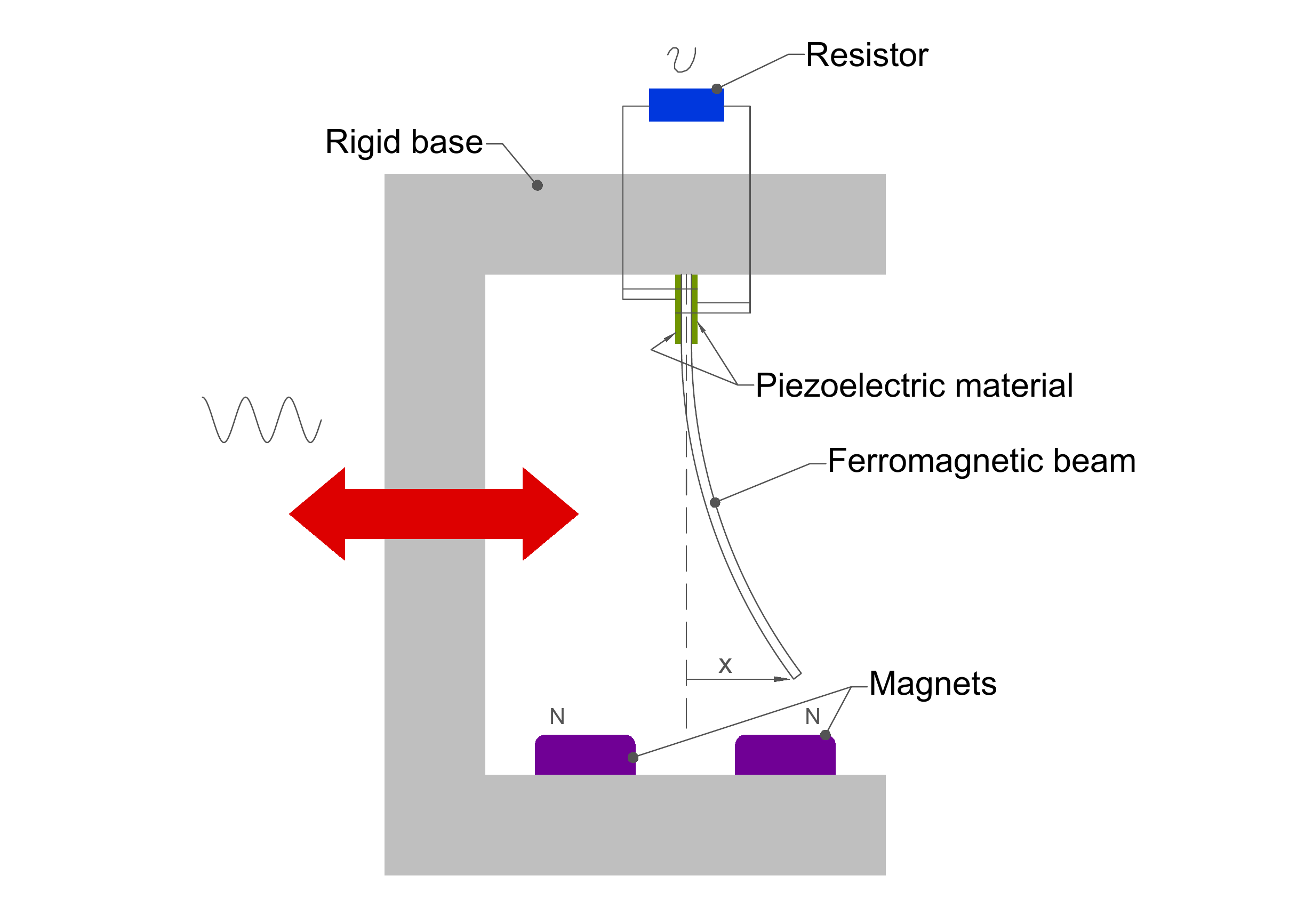}
\caption{Illustration of the bistable piezo-magneto-elastic energy harvesting system.}
\label{harvesting_device_fig}
\end{figure}

The energy harvesting system of interest in this work is the (sinusoidal excited) piezo-magneto-elastic beam presented by Erturk et al. \cite{erturk2009p254102}, which is based on the (stochastically excited)  inverted pendulum energy harvester proposed by Cottone et al. \cite{cottone2009p080601}. An illustration of this bistable energy harvesting system can be seen in Figure~\ref{harvesting_device_fig}, where it is possible to see that it consists of a vertical fixed-free beam made of ferromagnetic material, a rigid base and a pair of magnets. In the beam upper part, there is a pair of piezoelectric laminae coupled to a resistive circuit. The rigid base is periodically excited by a harmonic force, which, together with the magnetic force generated by magnets, induces large amplitude vibrations. The piezoelectric laminae convert the energy of movement 
into electrical power, which is dissipated in the resistor.

Although this system dissipates energy, instead of using it to supply some secondary system, it is a typical prototype of a piezoelectric energy harvesting system. In an application of interest, the resistor is replaced by a more complex electrical circuit, which stores (and possibly manipulates) the voltage delivered by the mechanical system.

\subsection{Initial value problem}

The dynamic behavior of the system of interest is described by the following initial value problem
\begin{equation}
	\ddot{x} + 2 \, \xi \, \dot{x} - \frac{1}{2} x \left(1-x^2\right) - \chi \, v
	= f \, \cos{\left( \Omega \, t \right)},
	\label{ode_mechanical}
\end{equation}
\begin{equation}
	\dot{v} + \lambda \, v + \kappa \, \dot{x} = 0,
	\label{ode_electrical}
\end{equation}
\begin{equation}
	x(0) = x_0, ~ \dot{x}(0) = \dot{x}_0, ~ v(0) = v_0,
	\label{initial_conditions}
\end{equation}
where $\xi$ is the damping ratio; $\chi$ is the piezoelectric coupling term in mechanical equation; $\lambda$ is a reciprocal time constant;  $\kappa$ is the piezoelectric coupling term in electrical equation; $f$ is the external excitation amplitude; $\Omega$ is the external excitation frequency. The initial conditions are $x_{0}$, $\dot{x}_{0}$ and $v_{0}$, which respectively represent, the beam edge initial position, initial velocity and the initial voltage over the resistor. Also, $t$ denotes the time, so that the beam edge displacement at time $t$ is given by $x(t)$, and the resistor voltage at $t$ is represented by $v(t)$. The upper dot is an abbreviation for time-derivative, and all of these parameters are dimensionless.

\subsection{Mean output power}

Since the objective in this work is to maximize the amount of energy recovered by an energy harvesting process, the main quantity of interest (QoI) associated with the nonlinear dynamical system under analysis is the mean output power
\begin{equation}
P = \frac{1}{T_f-T_0} \int_{\tau=T_0}^{T_f} \lambda \, v^2(\tau) \, d\tau,
\end{equation}
which is defined as the temporal average of the instantaneous power $\lambda \, v^2$ over a given time interval $[T_0,T_f]$. This QoI plays the role of objective function in the optimization problem defined in section~\ref{nonlinear_opt_prob}.

\subsection{Dynamic classifier}
\label{dyn_sys_classifier}

\begin{figure}[h]
\centering
\subfigure[regular dynamics]{\includegraphics[scale=0.3]{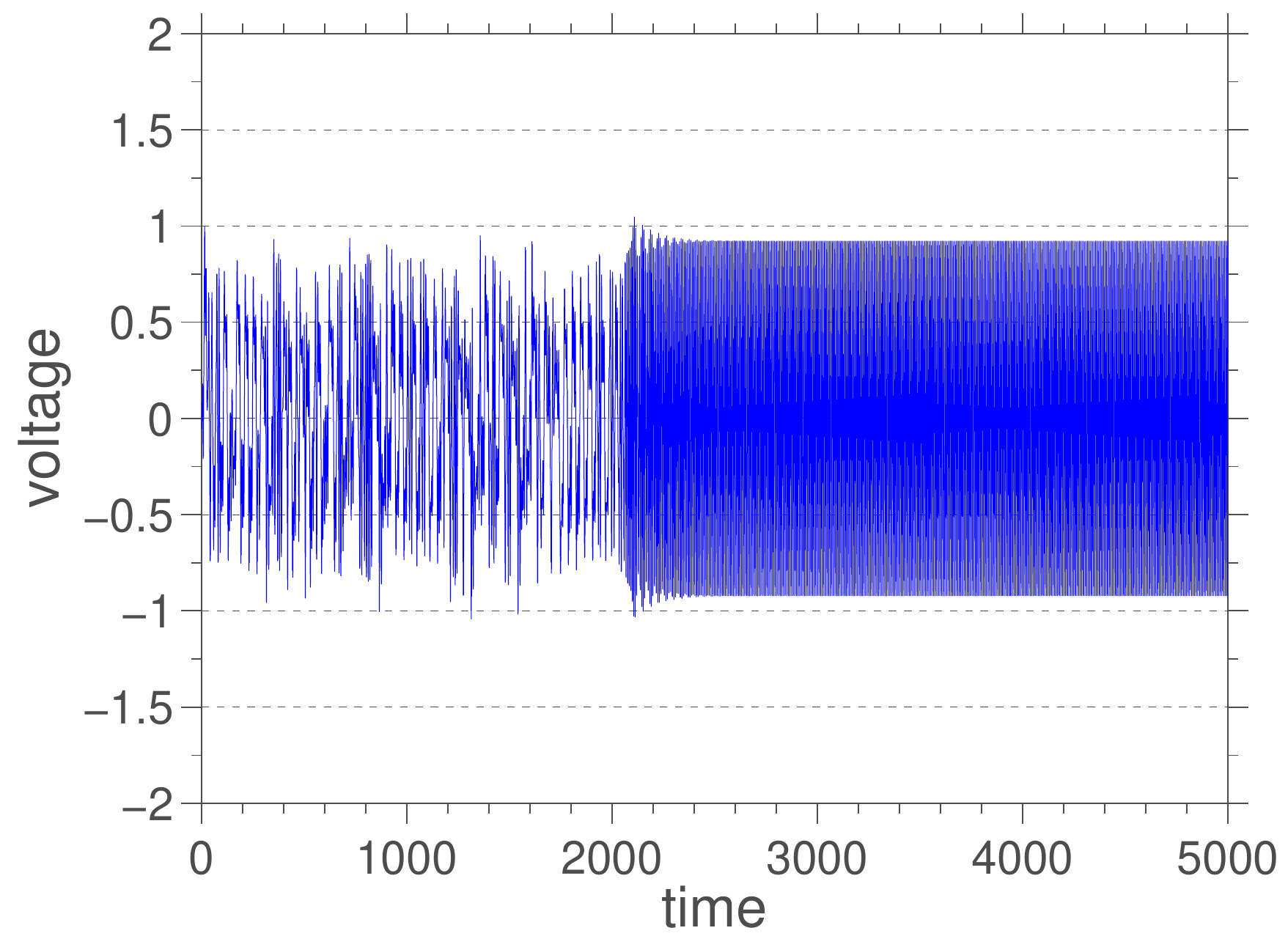}}
\subfigure[chaotic dynamics]{\includegraphics[scale=0.3]{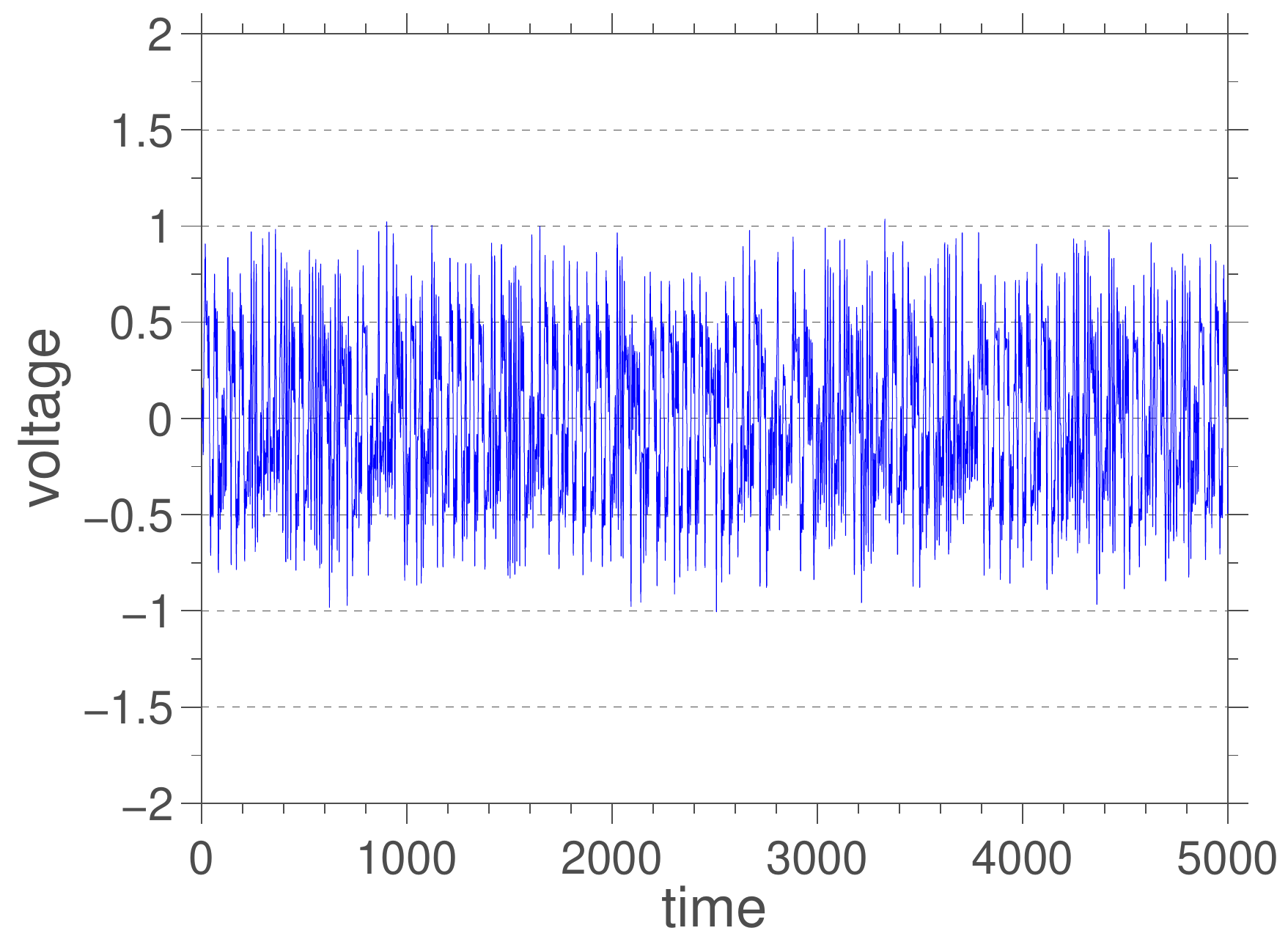}}
\caption{Two typical voltage time-series for the bistable energy harvesting system. The time-series in (a) has a regular steady-state dynamics, while in (b) a chaotic steady-state dynamics is observed.}
\label{voltage_series_fig}
\end{figure}

Due to dynamical system nonlinearity, the steady-state dynamical response (over a given time interval) of the energy harvesting device may be chaotic or regular (non-chaotic), such as illustrated in Figure~\ref{voltage_series_fig}, which shows typical voltage time-series for this kind of bistable oscillator.

To distinguish between the chaotic and regular dynamic regimes, the 0-1 test for chaos by  Gottwald and Melbourne \cite{gottwald2004p603,gottwald2009p129,gottwald2009p1367,gottwald2016} is employed. This test, which is based on an extension of the dynamical system to a two-dimensional Euclidean group \cite{litak2016p1433}, uses a binary classifier $K$ to identify the dynamic regime of operation underlying the system of interest. This classifier is constructed from a dynamical system observation (time-series) $\phi(t)$. In fact, it is sufficient to use $\bm{\Phi} = \left(\phi(t_1), \phi(t_2), \cdots, \phi(t_N)\right)$, a discrete version of the observable $\phi(t)$ that is obtained through a numerical integration (sampling) process.

Receiving the discrete observation (time-series) as an input, the analytical procedure of the 0-1 test consists of the following steps:
\begin{enumerate}
\item A real parameter $c \in [0, 2 \pi)$ is chosen and, for $n=1, 2, \cdots, N$, the discrete version of $\phi(t)$ is used to define the translation variables
\begin{equation}
	p_n (c) = \sum_{j=1}^{n} \phi(t_j) \, \cos{(j\,c)},
	\label{test01_pq1}
\end{equation}
\begin{equation}
	q_n (c) = \sum_{j=1}^{n} \phi(t_j) \, \sin{(j\,c)}.
	\label{test01_pq2}
\end{equation}
\item Then, for $n=1, 2, \cdots, N$, the time-averaged mean square displacement of the dynamics trajectory in $(p_c, q_c)$ space is computed by
\begin{equation}
	M_n (c) = \lim_{N \to \infty} \frac{1}{N} \, \sum_{j=1}^{N} \tilde{M}_n (c),
	\label{test01_MSD}
\end{equation}
where
\begin{equation}
	\tilde{M}_n (c) = \left(p_{j+n}(c)-p_{j}(c) \right)^2 + \left( q_{j+n}(c)-q_{j}(c) \right)^2.
\end{equation}

\item Finally, defining the mean-square vector 
\begin{equation}
	\vec{M}_n = \left( M_1, M_2, \cdots, M_n\right),
\end{equation}
and the temporal mesh vector
\begin{equation}
	\vec{t}_n = \left( t_1, t_2, \cdots, t_n \right),
\end{equation}
the dynamic classifier is constructed through the correlation
\begin{equation}
	K_c = \lim_{n \to \infty} \frac{ \cov{\vec{t}_n}{\vec{M}_n} }{ \sqrt{\varvar{\vec{t}_n} \, \varvar{\vec{M}_n}} },
	\label{test01_Kc}
\end{equation}
\noindent
where $\cov{\cdot}{\cdot}$ and $\varvar{\cdot}$ respectively denote the covariance 
and variance statistical operators.
\end{enumerate}

It can be proved that $K_c \in \{0,1\}$, being $K_c = 0$ for regular dynamics and $K_c = 1$ for chaotic dynamics \cite{gottwald2009p1367,litak2016p1433}. For numerical implementation purposes, the above procedure is performed several times, for several randomly chosen values of $c \in [0, 2\pi)$, and considering the voltage time-series as the system observation, i.e., $\phi(t) = v(t)$.  The discrete version of $\phi(t)$ is constructed through  temporal integration via the standard 4th order Runge-Kuta method. The limit processes in Eqs.(\ref{test01_MSD}) and (\ref{test01_Kc}) are replaced by the condition $n \ll N$, and the classifier $K$ is calculated as the median of $K_c$ realizations, i.e.,  $K = \median{K_c}$. Indeed, for a careful done numerical simulation one has $K \approx 0$ or $K \approx 1$. See \cite{litak2016p1433,gottwald2016} for further details.


\section{Non-convex optimization problem}
\label{nonlinear_opt_prob}

\subsection{Problem definition}

The objective of this work is to find a set of parameters that maximize the mean output power dissipated in the resistor. For this purpose, the electromechanical system properties ($\xi$, $\chi$, $\lambda$ and $\kappa$) sound as natural choices for the design variables, once they are intrinsic to the energy harvester embedded physical characteristics, something a designer might want to modify for better performance. However, the present work uses, first, the dynamical system excitation parameters, namely the amplitude $f$ and the frequency $\Omega$, as design variables. Although this choice sounds quite artificial at first since, in general, the designer does not have control of the level (or frequency) of vibration in which the energy harvester is operating, it is also of interest in the optimization of these systems to know in which vibration settings the device can recover more energy. As the numerical methodology developed here is completely general, it can be easily applied in the cases where the other parameters are used as the design variables, but by using $f$ and $\Omega$ to test the proposed methodology, an advantage is gained in terms of physical intuition about underlying physics, as the literature presents better knowledge about the effect of these two parameters on the behavior of this bistable energy harvesting system.

Many combinations of $(f,\Omega)$ lead the dynamical system to operate in the chaotic regime, an undesirable condition for electric power usage in principle, since the process of rectifying the chaotic electrical signal, which is necessary to enable its use in applications, can consume a significant part of the available energy. Although it is possible to explore the chaotic dynamics in favor of greater efficiency of the system \cite{Savi2015p2787,Daqaq2020p1525}, as done by the author and collaborators in \cite{cunhajr_cobem2019_1}, through the use of chaos control techniques, this is not the approach followed in the present work, which seeks to develop a very general numerical framework.

Thus, as not every pair $(f,\Omega)$ is an acceptable choice for the optimal design, it is necessary to impose a constraint that ensures the regularity (non-chaoticity) of the system dynamics. Taking advantage of the 0-1 test for chaos, described in section~\ref{dyn_sys_classifier}, the constraint to ensure a regular (non-chaotic) dynamic regime can be formulated as $K = 0$. Note that the optimization problem is extremely nontrivial, once the constraint presents jump-type discontinuities since $K\in\{0,1\}$.

\subsection{Constrained problem formulation}

In an abstract way, one can formulate the constrained nonlinear optimization problem described above as find a feasible vector of design variables $\vec{x}^{\star}$ that maximize a certain objective function $\vec{x} \in \mathcal{D} \mapsto \mathcal{S}(\vec{x}) \in \R$, i.e.,
\begin{equation}
\vec{x}^{\star} = \arg\max_{\vec{x} \in \mathcal{D}}{ \, \mathcal{S}(\vec{x})},
\label{opt_prob_constr_eq}
\end{equation}
\noindent
where the set of admissible (feasible) parameters is defined by the bounded region
\begin{equation}
	\mathcal{D} = \left\lbrace \vec{x}
	\,\,\vert\,\, 
	\vec{x}_{min} \leq \vec{x} \leq \vec{x}_{max}
	~~ \mbox{and} ~~
	\mathcal{G}(\vec{x}) = 0 \right\rbrace.
\end{equation}

In this context, it is straightforward to see that the design variables vector is $\vec{x} =(f,\Omega)$, the objective function is $\mathcal{S}(\vec{x}) = P(f,\Omega)$, the binary constraint is $\mathcal{G}(\vec{x})=K(f,\Omega)$, and the limits of the design variables are $\vec{x}_{min}=(f_{min},\Omega_{min})$ and $\vec{x}_{max}=(f_{max},\Omega_{max})$. 

In the case of a problem with more variables, or with different objective function and/or constraints, the adaptations to be made are straightforward.

\subsection{Penalized problem formulation}
\label{penal_opt_prob}

A penalized version of this constrained nonlinear optimization problem is introduced here in order to facilitate the computational implementation of the solution algorithm. In this formulation the constraint $\mathcal{G}(\vec{x}) = 0$ is replaced by the weaker condition $\mathcal{G}(\vec{x}) \leq \varepsilon \ll 1$, once in practice the best one has is $K \approx 0$. In this way, the problem defined by (\ref{opt_prob_constr_eq}) is replaced by the penalized problem which seeks a pair $\vec{x}^{\star}$ such that
\begin{equation}
\vec{x}^{\star} = \arg\max_{\vec{x} \in \mathcal{D}}{\tilde{\mathcal{S}}(\vec{x})},
\label{opt_prob_penal_eq}
\end{equation}
\noindent
where the set of feasible parameters is now defined by
\begin{equation}
	\mathcal{D} = \left\lbrace \vec{x}
	\,\,\vert\,\, 
	\vec{x}_{min} \leq \vec{x} \leq \vec{x}_{max} \right\rbrace,
\end{equation}
and the penalized objective function is given by
\begin{equation}
\tilde{\mathcal{S}}(\vec{x}) =  \mathcal{S}(\vec{x}) \,-\, \alpha \, \max \left\lbrace 0, \mathcal{G}(\vec{x}) \,-\, \varepsilon \right\rbrace.
\end{equation}

\noindent
The penalty parameter is heuristically chosen, being $\alpha = 10$ the value used in all simulations reported in this work. In the same way, $\varepsilon = 1/10$ is adopted.

\section{The cross-entropy method}
\label{ce_method}

\subsection{The general idea of CE method}

The key idea of the CE method is to transform the given non-convex optimization problem into an ``equivalent'' rare event estimation problem, that can be efficiently treated by a Monte Carlo like algorithm. The only requirement is that the problem has a single solution, i.e., the global extreme is unique.

In this framework, the feasible region is sampled according to a given probability distribution chosen by the user, and the low-order statistics (mean and standard deviation) of these samples are used to update the optimum point estimation, and the stopping criteria metric (respectively). This is a two-step iterative process:

\begin{enumerate}
	\item \emph{Sampling}: First, the feasible region is sampled according to a given probability distribution, and then the objective function is evaluated in each one of these samples. The information embedded in these samples are used in the algorithm adaptive process;
	\item \emph{Learning}: A special subset of these samples, dubbed the \emph{elite sample set}, is defined by the samples that produced the highest values for the objective function. The parameters of the probability distribution are updated using statistics obtained from this elite sample set, modifying the given distribution in a sense that tries to make it as close as possible to a Dirac delta centered on the global optimum. 
\end{enumerate}

Throughout this process, the distribution mean provides the approximation for the global optimum. Its update is done in the sense of moving the distribution center towards the optimization problem optimum, while the standard deviation is reduced, thus ``shrinking'' the distribution around its central value. The continuous composition of these effects of translation and ``shrinking'' characterizes the process in which the distribution is ``shaped'' towards a point mass Dirac distribution centered on the optimal.

\subsection{Theoretical framework}

Without loss of generality, suppose that the problem of interest is to maximize an objective function $\vec{x} \in \mathcal{D} \mapsto \mathcal{S}(\vec{x}) \in \R$, as stated in Eqs.(\ref{opt_prob_constr_eq}). If the penalized formulation from Eq.(\ref{opt_prob_penal_eq}) is adopted, just consider $\vec{x} \in \mathcal{D} \mapsto \tilde{\mathcal{S}}(\vec{x}) \in \R$ instead of $\mathcal{S}(\vec{x})$.

Denote the global optimal by $\vec{x}^{\star}$ and the corresponding maximum value by $\gamma^{\star} = \mathcal{S}(\vec{x}^{\star})$, i.e., $\gamma^{\star} = \max{ \, \mathcal{S}(\vec{x})}$ for all $\vec{x} \in \mathcal{D}$.

Using this global maximum as a reference value, and considering a randomized version of the vector $\vec{x}$, denoted by $\vec{X}$, it is possible construct the random event $\mathcal{S}(\vec{X}) \geq \gamma^{\star}$, which represents the scenarios where the random variable $\mathcal{S}(\vec{X})$ is bigger or equal to the (deterministic) scalar value $\gamma^{\star}$.  In other words, this random event considers the possibility of choosing, randomly, points in the feasible region $\mathcal{D}$ that produce values greater than or equal to the global maximum. Note that this random event is related to the optimization problem defined in Eq.(\ref{opt_prob_constr_eq}), it can be thought of as its randomized version.

Since $\gamma^{\star}$ is the maximum value of $\mathcal{S}(\vec{x})$, no realization of $\vec{X}$ can produce $\mathcal{S}(\vec{X}) > \gamma^{\star}$ and only $\vec{X}=\vec{x}^{\star}$ can make $\mathcal{S}(\vec{X}) = \gamma^{\star}$. Therefore, the probability of this random event is zero, i.e., $\PM \left\lbrace \mathcal{S}(\vec{X}) \geq \gamma^{\star} \right\rbrace  = 0$.

Relaxing the reference value for a scalar $\gamma < \gamma^{\star}$, one has the random event $\mathcal{S}(\vec{X}) \geq \gamma$, which defines a rare event if $\gamma \approx \gamma^{\star}$, for which $\PM \left\lbrace \mathcal{S}(\vec{X}) \geq \gamma \right\rbrace  \approx 0$.

In his 1997 seminal work \cite{Rubinstein1997p89}, R. Rubinstein conceived the CE method as a tool to efficiently estimate such type of probabilities,  associated with events located at the distribution tail.  Sometime later \cite{Rubinstein1999p127}, due to the connection between the random event and the optimization problem described above, he noticed that this rare event estimation process could be used as a global search method to optimize the given objective function.

Notice that, with the aid of the expected value operator $\expval{\cdot}$, and the indicator function
\begin{equation}
	\indfunc{\mathcal{A}}(\vec{x}) =  
	\begin{cases} 
		1, ~ \mbox{if} ~ \vec{x} \in \mathcal{A} \\ 
		0, ~ \mbox{if} ~ \vec{x} \notin \mathcal{A} \, , \\ 
		\end{cases}
\end{equation}
the probability of the random event $\mathcal{S}(\vec{X}) \geq \gamma$ can be written as
\begin{equation}
\PM \left\lbrace \mathcal{S}(\vec{X}) \geq \gamma \right\rbrace =
\expval{\indfunc{\mathcal{S}(\vec{X}) \geq \gamma}},
\label{eq_prob_ce}
\end{equation}
which provides a practical way of estimating its value. Once the right side of Eq.(\ref{eq_prob_ce}) is the mean value of the random event $\indfunc{\mathcal{S}(\vec{X}) \geq \gamma}$, it can be easily computed through the sample mean of $N_s$ samples of $\vec{X}$, i.e.,
\begin{equation}
\expval{\indfunc{\mathcal{S}(\vec{X}) \geq \gamma}} \approx
\frac{1}{N_s} \, \displaystyle \sum_{k=0}^{N_s} \indfunc{\mathcal{S}(\vec{X}_k) \geq \gamma} \, ,
\end{equation}
where the realizations $\vec{X}_k$ ($k=1, \cdots, N_s$) are drawn according to $\pdf{} \, (\vec{x}; \vec{v})$, the probability density function of $\vec{X}$, parametrized by the hyper-parameters vector $\vec{v}$. Very often, $\vec{v} = (\bm \mu, \bm \sigma)$, where $\bm \mu$ and $ \bm \sigma$ are the the mean the standard deviation vectors of $\vec{X}$, respectively.

The idea of the CE method is to approximate the solution of the underlying optimization problem by solving the rare event probability estimation from Eq.(\ref{eq_prob_ce}). For this purpose, it employs a multilevel approach, that generates an optimal sequence of statistical estimators for the pair $(\gamma,\vec{v})$, denoted by $\left( \estim{\gamma}_{\ell}, \estim{\vec{v}}_{\ell} \right)$, such that
\begin{equation}
\estim{\gamma}_{\ell} \xrightarrow{~a.s.~} \gamma^{\star}
~~ \mbox{and} ~~ 
\pdf{} \, (\vec{x},\estim{\vec{v}}_{\ell}) \xrightarrow{~a.s.~} \delta \left(\vec{x} - \vec{x}^{\star} \right),
\end{equation}
i.e., the reference level $\gamma$ tends with probability 1 to the maximum value, and the family of distributions $\pdf{} (\cdot \,,\,\vec{v})$ goes (almost sure) towards a point mass distribution, centered on the optimization problem global optimum. 

This sequence of estimators is optimal in the sense that it minimizes the Kullback-Leibner divergence between $\indfunc{\mathcal{S}(\vec{X}) \geq \gamma}$ and $\pdf{} (\cdot \,,\,\vec{v})$ \cite{Rubinstein2016}.

More concretely, the feasible region $\mathcal{D}$ is sampled with $N_s$  independent and identically distributed (iid) realizations of the random vector $\vec{X}$, drawn from its density $\pdf{} \, (\vec{x}; \vec{v})$. For each of these samples, the objective function is evaluated, generating the sequence of values $\mathcal{S}(\vec{X}_1), \mathcal{S}(\vec{X}_2), \cdots, \mathcal{S}(\vec{X}_{N_s})$. 

Then, an elite sample $\mathcal{E}_t = \left\lbrace \vec{X}_k : \mathcal{S}(\vec{X}_k) \geq \estim{\gamma}_t \right\rbrace$
is defined by lumping the $N_e < N_s$ points that better performed, i.e., those which produced the highest values for $\mathcal{S}(\vec{x})$.
Note that this elite set is defined in terms of the maximum value statistical estimator, given by
\begin{equation}
	\estim{\gamma}_t = \mathcal{S}_{(N_s-N_e+1)}.
	\label{orderstat_gamma_eq}
\end{equation}

The hyper-parameters vector $\vec{v}$ is updated using the maximum likelihood estimator so that, with the aid of the elite set $\mathcal{E}_t$, it is written as 
\begin{equation}
	\estim{\vec{v}}_t =
	\arg\,\max_{\vec{v}} \,\sum_{\vec{X}_k \in \mathcal{E}_t} \ln \left( \pdf{} \, (\vec{X}_k; \vec{v}) \right).
	\label{MLE_v_eq}
\end{equation}

Depending on the distribution chosen for $\vec{X}$, the stochastic program from Eq.(\ref{MLE_v_eq}) needs to be solved numerically. However, for the distributions of the exponential family, which includes the Gaussian and its truncated version, this estimator can be calculated in a analytic way, with each component of mean and deviation given by the formulas
\begin{equation}
	\estim{\mu}_t = \frac{\displaystyle \sum_{\vec{X}_k \in \mathcal{E}_t} X_{k}}{N_{e}},
\end{equation}
and
\begin{equation}
\estim{\sigma}_t = \sqrt{\frac{\displaystyle \sum_{\vec{X}_k \in \mathcal{E}_t} \left(X_{k} - \estim{\mu}_t\right)^2}{N_{e}}},
\end{equation}
respectively.

Although this process has a theoretical guarantee of converging to a point mass distribution centered on the global optimum, in computational practice it is common to see the updated distribution numerically degenerating before it ``reaches the target''. Sometimes the standard deviation decreases very quickly, causing the distribution to ``shrinks''  in a region far from the global optimum. In this scenario, only samples far from the optimum point are drawn, producing poor estimates for the optimization problem solution.

At the theoretical level, where it is possible to sample an infinite number of times, this pathological situation is bypassed, because at some moment a sample on the tail is drawn, moving the distribution center out of the ``frozen region'', as this outlier has a lot of weight in the mean estimate. However, in computational terms, as any robust implementation requires a maximum number of iterations (levels), the algorithm may stop with the distribution center within a ``frozen region''.

This problem may be solved through the use of a \emph{smooth updating scheme} for the hyper-parameters
\begin{equation}
\estim{\bm \mu}_t := \alpha \, \estim{\bm \mu}_t + (1-\alpha) \, \estim{\bm \mu}_{t-1},
\end{equation}
\begin{equation}
\estim{\bm \sigma}_t := \beta_t \, \estim{\bm \sigma}_t + (1-\beta_t) \, \estim{\bm \sigma}_{t-1},
\end{equation}
\begin{equation}
\beta_t  =  \beta - \beta \, \left(1 - \frac{1}{t} \right)^q,
\end{equation}
where the smooth parameters are such that $0 < \alpha \leq 1$, $0.8 \leq \beta \leq 0.99$ and $5 \leq q \leq 10$ \cite{kroese2011,Rubinstein2004}, and the estimations at $t$ and $t-1$ are obtained by solving the Eq.(\ref{MLE_v_eq}), which defines the nonlinear program that gives rise to the vector $\vec{v}$.

\subsection{The computational algorithm}
The geometric idea of the CE method is described at the beginning of this section, being complemented by the theoretical formalism presented in the previous subsection. The compilation of these ideas in the form of an easy to implement computational algorithm is presented below:

\begin{enumerate}
\item Define the number of samples $N_{s}$, the number of elite samples $N_{e} < N_{s}$, a convergence tolerance $\texttt{tol}$, the maximum of iteration levels $\ell_{max}$, a family of probability distributions $\pdf{} \left( \cdot, \vec{v}{} \right)$, an initial vector of hyper-parameters $\estim{\vec{v}}_0$ for $\pdf{}$, and set the level counter $\ell=0$; 

\item Update the level counter $\ell = \ell+1$;

\item Generate a total of $N_{s}$ independent and identically distributed (iid) samples from $\pdf{} \left( \cdot, \estim{\vec{v}}_{\ell-1} \right)$, denoted by $\vec{X}_1, \cdots, \vec{X}_{N_s}$;

\item Evaluate the objective function $\mathcal{S}(\vec{x})$ at the samples $\vec{X}_1, \cdots, \vec{X}_{N_s}$, sort the results $\mathcal{S}_{(1)} \leq \cdots \leq \mathcal{S}_{(N_s)}$, and define the elite sample set $\mathcal{E}_t$ with the $N_{e}$ points which better performed;

\item Update the estimators $\estim{\gamma}_{\ell}$ and $\estim{\vec{v}}_{\ell}$ with aid of the elite sample set, 
using order and maximum likelihood statistic estimators, given by the Eqs.(\ref{orderstat_gamma_eq}) and (\ref{MLE_v_eq}), respectively. If necessary, apply the scheme of smooth updating;

\item Repeat the steps (2) up to (5) of this algorithm while a (standard deviation dependent) stopping criterion is not met. For instance, $\max \left\lbrace \bm \sigma \right\rbrace < \texttt{tol}$.
\end{enumerate}

\begin{figure*}
\centering
\includegraphics[scale=0.5]{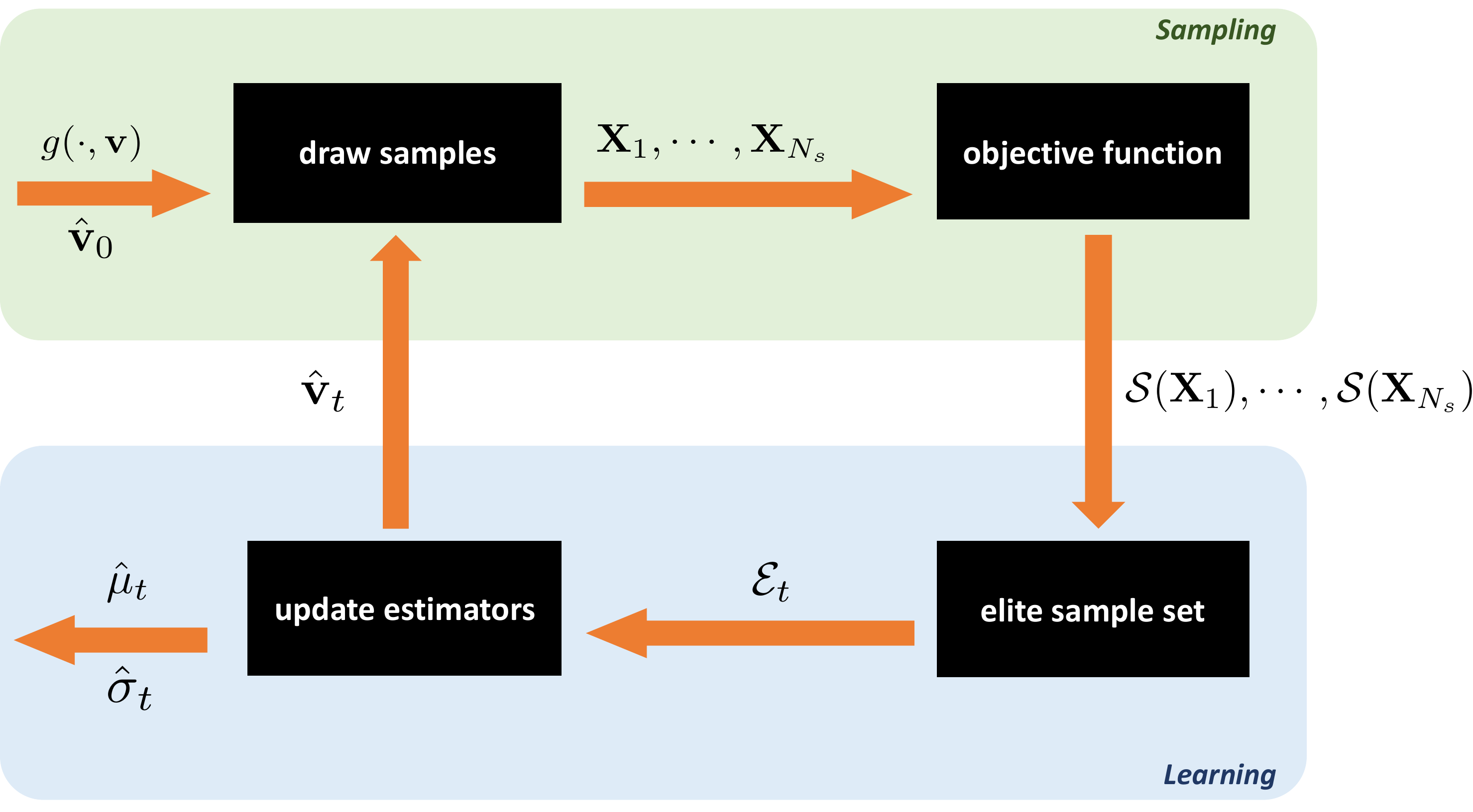}
\caption{Schematic representation of the CE algorithm for optimization.}
\label{ce-diagram_fig}
\end{figure*}

A schematic representation of this algorithm, illustrating all the stages of the sampling and learning phases, can be seen in Figure~\ref{ce-diagram_fig}.

\subsection{Remarks about CE method}

Among the several  characteristics that make this relatively novel technique interesting, the following can be highlighted:

\begin{itemize}
	\item \emph{Simplicity} -- very intuitive algorithm with few control parameters ($N_s$, $N_e$, $\ell_{max}$ and $\texttt{tol}$), each of then with a very clear interpretation; 
	\item \emph{Robustness} -- theoretical results ensure that, under typical conditions, the method is guaranteed to converge if the problem has a single global extreme;
	\item \emph{Efficiency} -- the method typically presents fast convergence in comparison with classic global search meta-heuristics, such as genetic algorithms; 
	\item \emph{Generality} -- the method does not require any regularity of the objective function and can be applied to almost any type of non-convex optimization problem (even non-differentiable or discontinuous);
	\item \emph{Extensibility} -- the theory is general, in principle it can be applied to problems of any finite dimension, the computational cost and the ``curse of dimensionality'' being the limiting factors in practice;
	\item \emph{Easy implementation} -- the algorithm can be implemented with a few lines of code in a high level programming language.
\end{itemize}

The mathematical development associated with the formalism described above is relatively non-trivial, but it has been well established throughout the first decade of the method, including theorems that strictly establish the conditions where the method has guaranteed convergence. These details are suppressed from this paper as they are outside the scope of the journal. But for the interested reader, the following references are recommended \cite{DeBoer2005p19,kroese2011,Rubinstein2004,Rubinstein2016}.

The computational experiments in the next section illustrate the efficiency and robustness of this framework in solving a non-trivial problem of optimizing an energy harvesting device.

\section{Numerical experiments}
\label{num_experim}

For the numerical experiments conducted here, the following numerical values are adopted for the dynamical system parameters: $\xi=0.01$, $\chi=0.05$, $\kappa=0.5$ and $\lambda=0.05$. The initial condition is defined by $x_0 = 1$, $\dot{x}_0 = 0$ and $v=0$. The dynamics is integrated over the time interval $[0, 2500]$, and the mean output power is computed over the last 50\% of this time-series, i.e., $[T_0,T_f] = [1250, 2500]$.

\subsection{Reference solution}
\label{ref_solution}

\begin{figure}[h]
\centering
\subfigure[]{\includegraphics[scale=0.35]{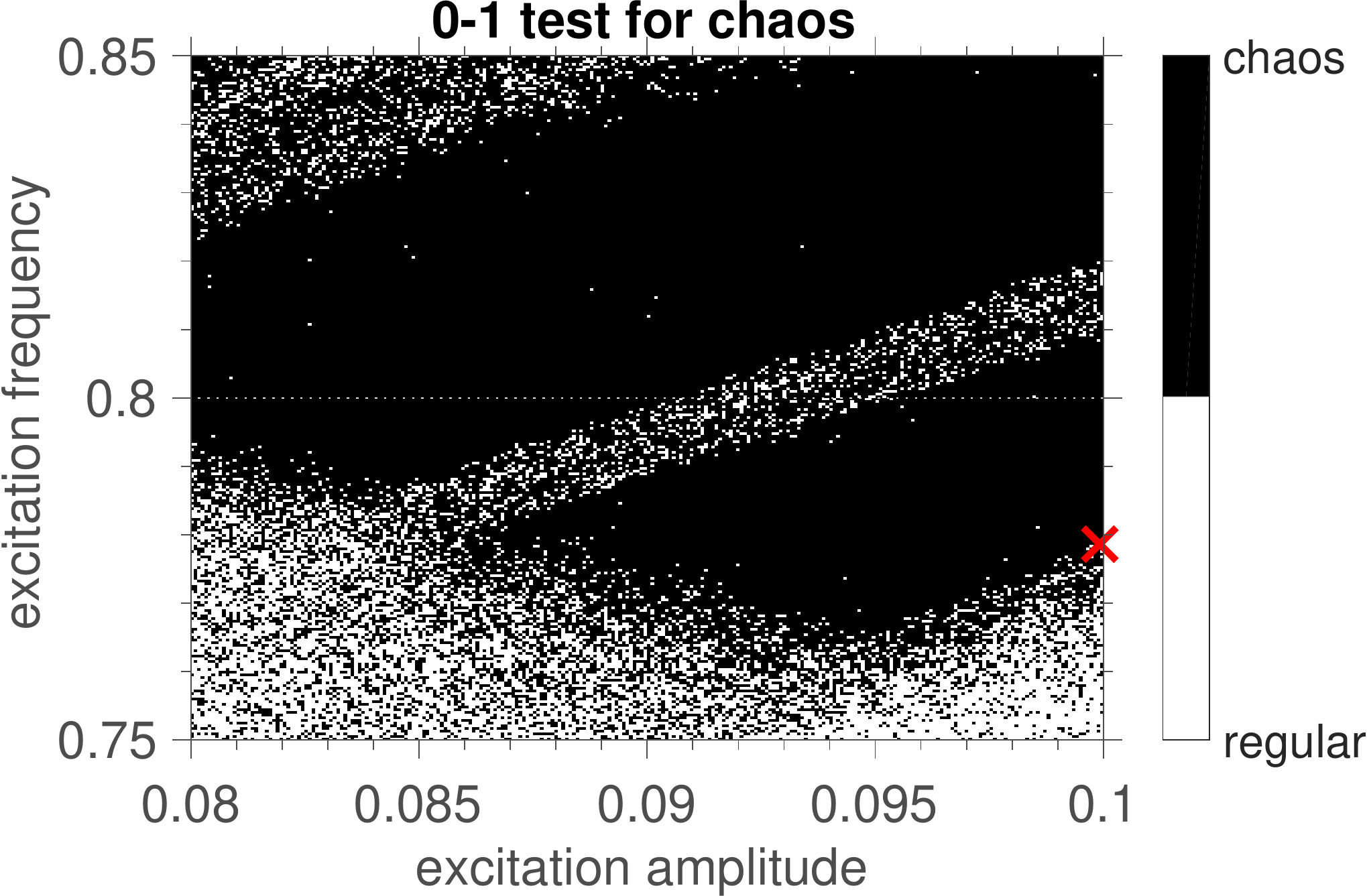}} ~~
\subfigure[]{\includegraphics[scale=0.35]{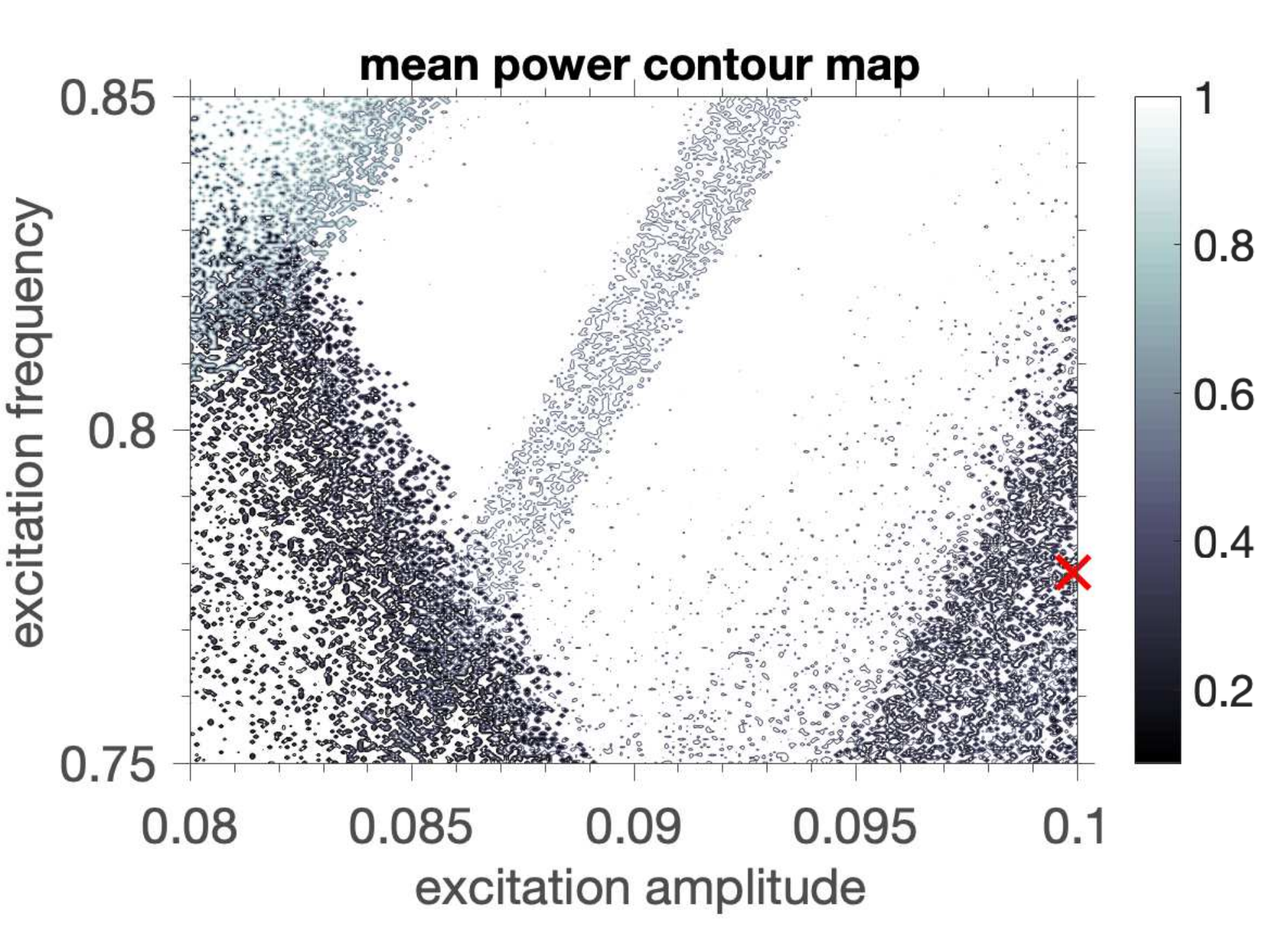}}
\caption{Bi-dimensional case contour maps: (a) constraint function defined by the 0-1 test for chaos, 
and (b) normalized objective function, defined by the mean output power.}
\label{ref_solution_fig}
\end{figure}

\begin{figure}[h]
\centering
\subfigure[]{\includegraphics[scale=0.7]{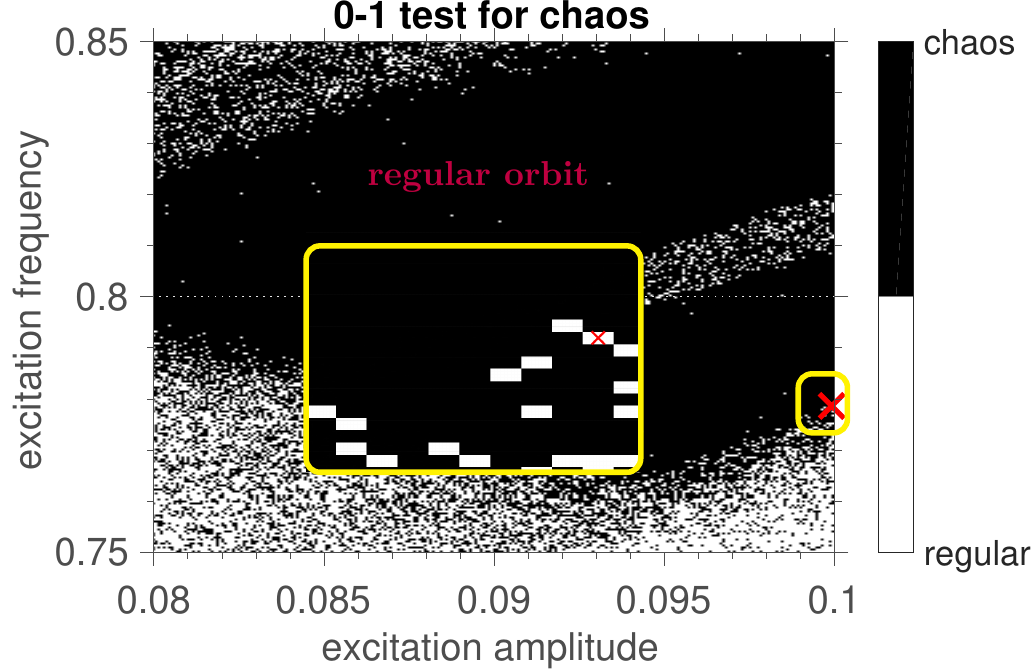}} ~~
\subfigure[]{\includegraphics[scale=0.7]{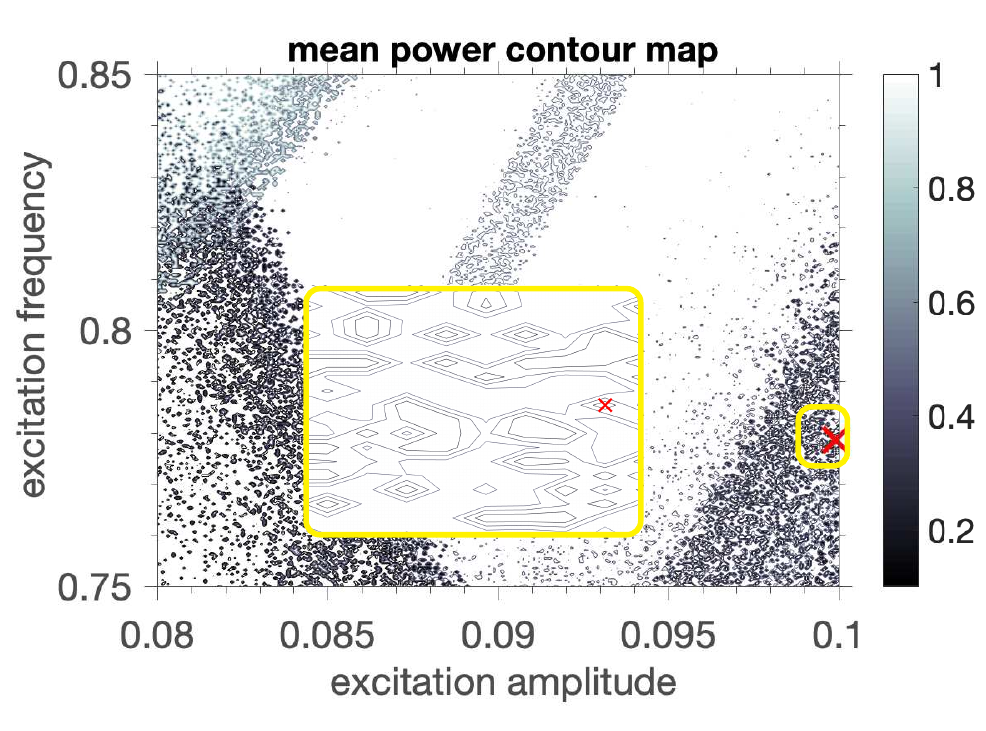}}
\caption{Magnification of the contour maps in bi-dimensional case: (a) constraint function defined by the 0-1 test for chaos, and (b) normalized objective function, defined by the mean output power.}
\label{ref_solution_fig2}
\end{figure}

In order to analyze the effectiveness and robustness of the stochastic solution strategy proposed here, a reference solution is computed by a standard exhaustive search on a fine grid over the domain $$\mathcal{D} = \left\lbrace (f,\Omega) \,|\, 0.08 \leq f \leq 0.1, \, 0.75 \leq \Omega \leq 0.85 \right\rbrace.$$
In this standard approach, a structured 256 x 256 uniform numerical grid is used to discretize $\mathcal{D}$. The system dynamics is then integrated for each grid point, with the optimization constraint being evaluated next. The objective function is evaluated at all feasible points, and the extreme value is updated at each step of the grid screening process. Two contour maps, associated with the reference solution, are shown in Figure~\ref{ref_solution_fig}: (a) constraint function, and (b) objective function normalized by $P_{max}^{DS}$. The pair $(f,\Omega) = (0.0999, 0.7786)$ that corresponds to the global maximum is indicated in both contour maps by a red cross, being associated with a mean output power $P_{max}^{DS} = 0.0172$. Figure~\ref{ref_solution_fig2} shows a magnification of the global maximum neighborhood, where it is possible to better appreciate the contour levels shape, and verify that it corresponds to a regular dynamic regime configuration. Note also that this result is compatible with the literature, which points to a better performance of piezoelectric harvesters for low frequencies and high amplitudes of excitation \cite{cunhajr_belhaq2019,Vocca2012p771}. For sake of reference, this solution was obtained in 3.6 hours in a \texttt{Dell Inspiron Core i7-3632QM 2.20 GHz RAM 12GB}.

\subsection{Cross-entropy solution}
\label{ce_num_result}

In the approach based on the CE method, the domain is randomly sampled using for this $N^{s} = 50$ points. This sampling is done according to a truncated Gaussian distribution,  parametrized by mean vector $\bm{\mu} = (\mu_{f}, \mu_{\Omega})$ and standard deviation vector $\bm{\sigma} = (\sigma_{f}, \sigma_{\Omega})$. The number of elite samples is chosen as $N^{e} = \texttt{round} (N^{s}/10)$, maximum number of levels is set $t_{max} = 100$, while the convergence criterion is adopted as $\max{ \left\lbrace \sigma_{f}, \sigma_{\Omega} \right\rbrace } < \texttt{tol}$, for a tolerance $\texttt{tol} = 1 \times 10^{-3}$. The smoothing parameters are $\alpha = 0.7$, $\beta = 0.8$ and $q=5$.

\begin{figure*}
\centering
\subfigure[ $\ell=1$]{\includegraphics[scale=0.37]{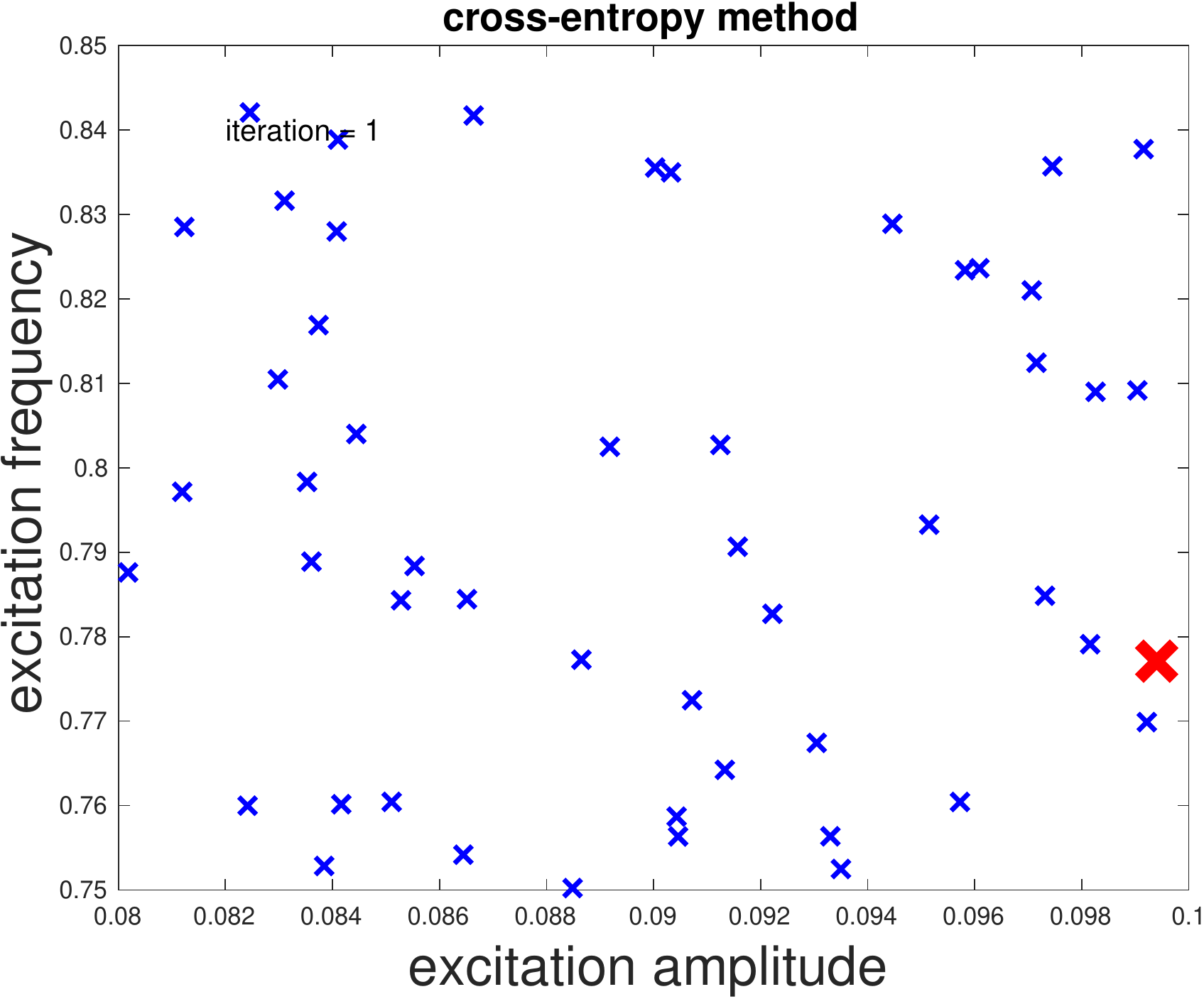}} ~~
\subfigure[ $\ell=5$]{\includegraphics[scale=0.37]{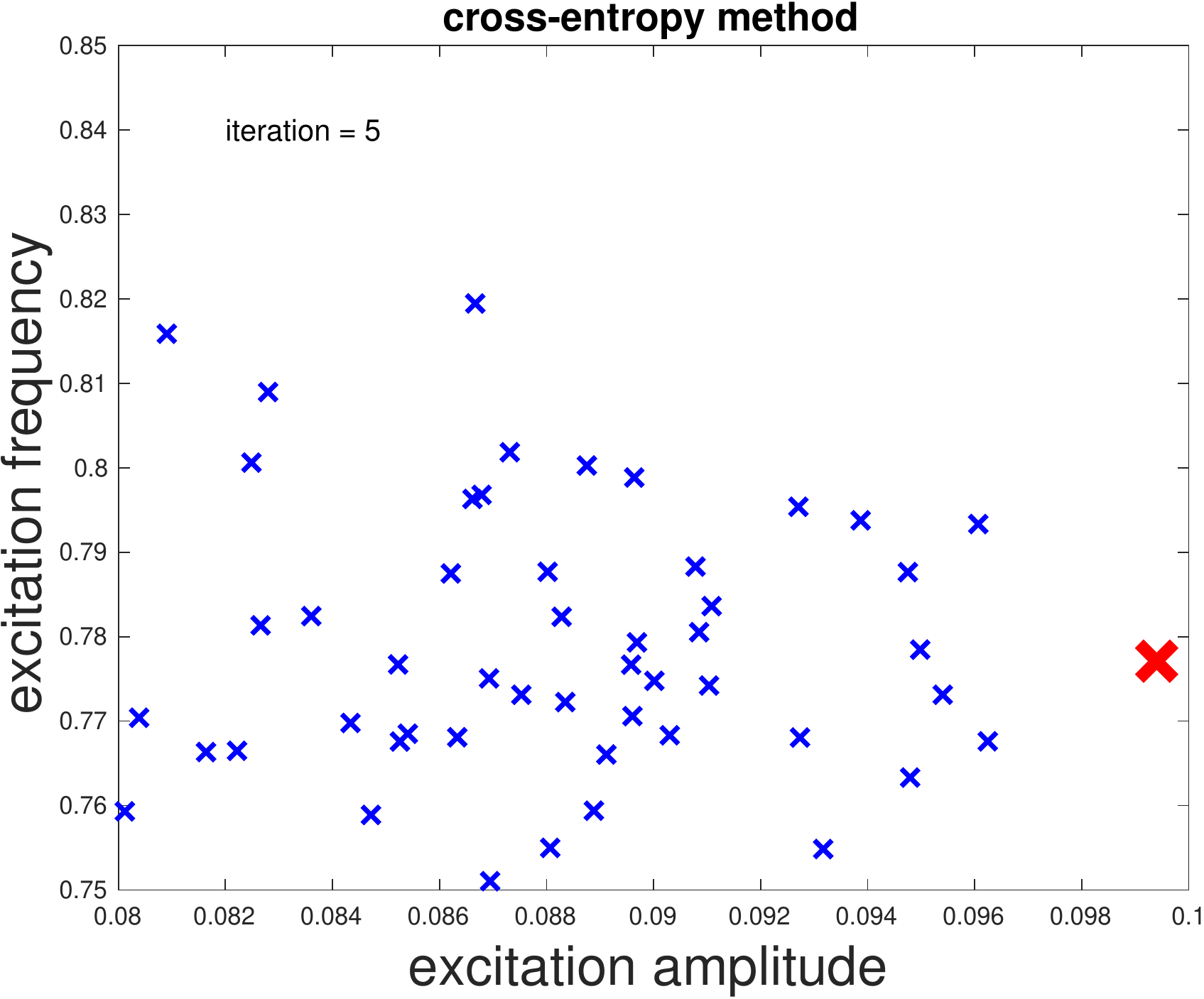}}\\
\subfigure[ $\ell=9$]{\includegraphics[scale=0.37]{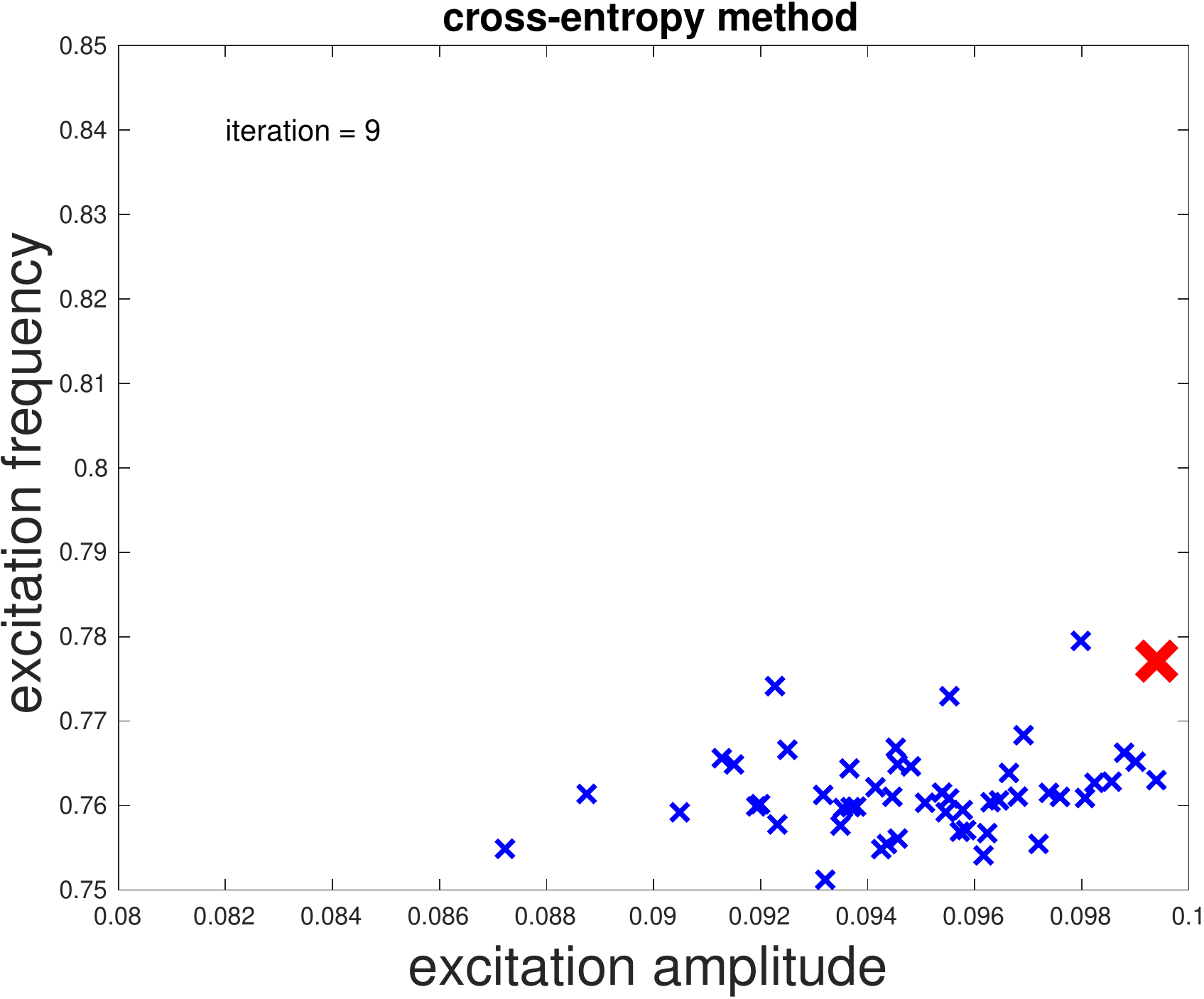}} ~~
\subfigure[$\ell=13$]{\includegraphics[scale=0.37]{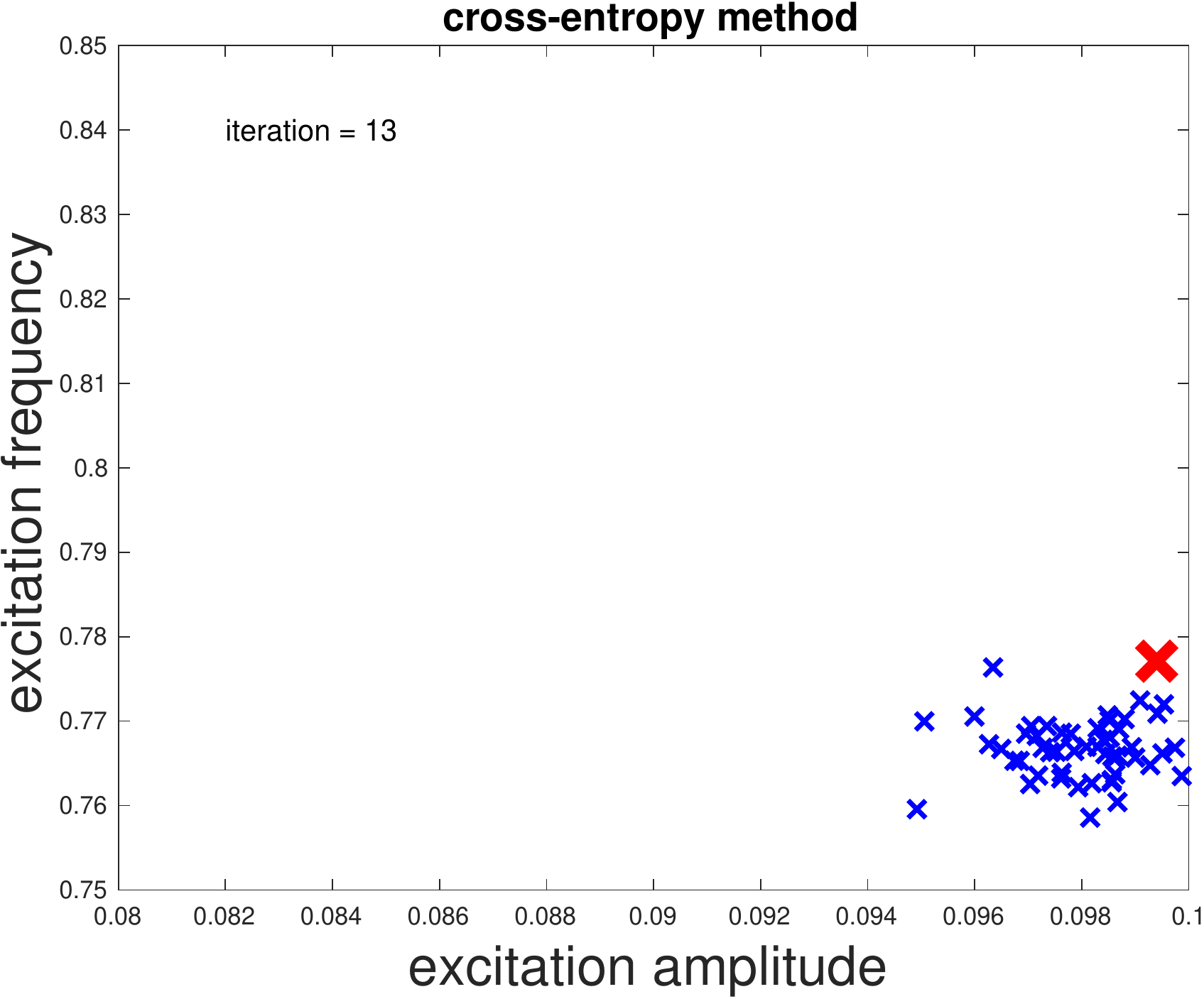}}\\
\subfigure[$\ell=20$]{\includegraphics[scale=0.37]{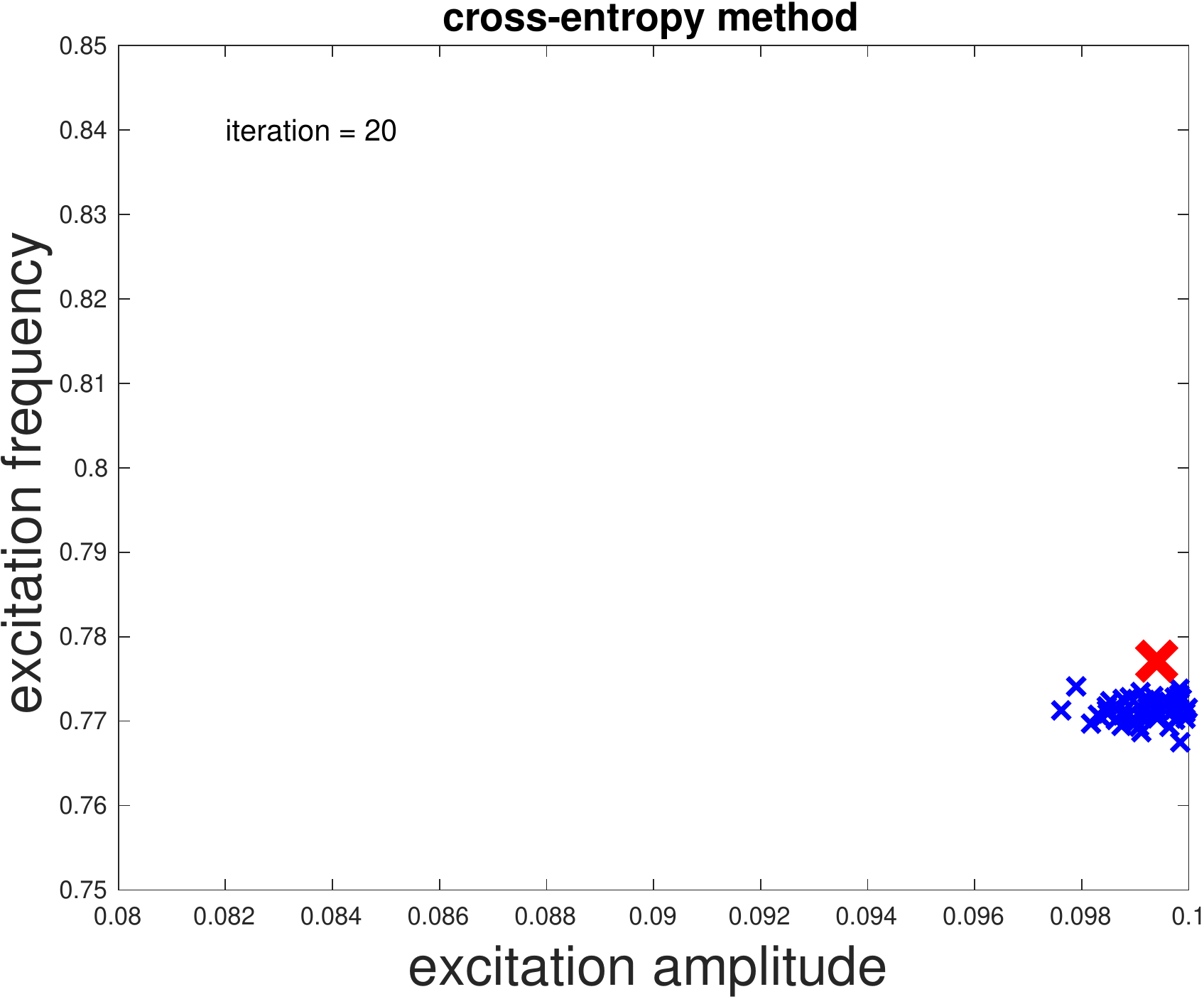}} ~~
\subfigure[$\ell=26$]{\includegraphics[scale=0.37]{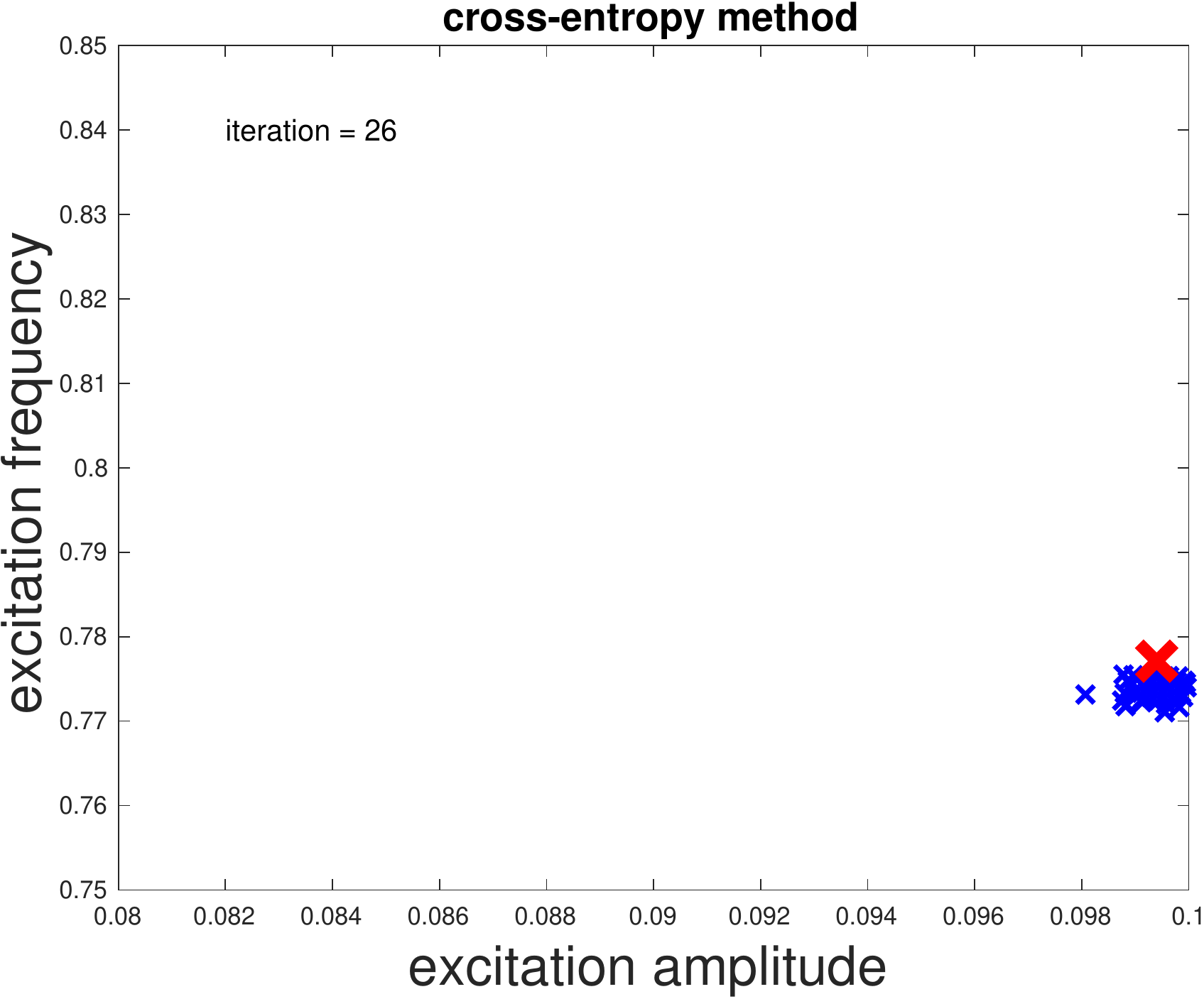}}\\
\caption{Illustration of CE method in the bi-dimensional case, using $N^{s} = 50$ samples, at different levels (iterations) of the algorithm. The reference solution is indicated with a red cross.}
\label{cross-entropy_fig50}
\end{figure*}

A visual illustration of the CE method is presented in Figure~\ref{cross-entropy_fig50}, which shows the domain sampling at different levels (iterations) of the algorithm. The reader can also appreciate the evolution of this random algorithm in Table~\ref{cross-entropy_tab50}, where each line displays the level index, the value of the means and standard deviations of $f$ and $\Omega$ in addition to the optimal value obtained for the objective function $P$ and the corresponding constraint $K$. In this case, the optimum value obtained by the CE method, with aid of the estimation from Eq.(\ref{orderstat_gamma_eq}), is $P_{max}^{CE} = 0.0170$, at the optimal point $\vec{x}^{\star} = (f^{\star},\Omega^{\star})$ where $f^{\star} \approx 0.10$ and $\Omega^{\star} \approx 0.77$.

\begin{table}[ht!]
\caption{Evolution of CE algorithm in the bi-dimensional case using $N^{s} = 50$ samples.}
\begin{tabular}{ccccccc}
\toprule
$\ell$ & $P$ & $K$ & $\mu_{f}$ & $\mu_{\Omega}$ &  $\sigma_{f}$ & $\sigma_{\Omega}$\\
\midrule
01 & 0.0151 & 0.1115 & 0.0858 & 0.7845 & 0.0260 & 0.1107 \\
02 & 0.0003 & 0.0195 & 0.0872 & 0.7804 & 0.0115 & 0.0498 \\
03 & 0.0004 & 0.0956 & 0.0901 & 0.7687 & 0.0084 & 0.0260 \\ 
04 & 0.0003 & 0.0610 & 0.0913 & 0.7674 & 0.0061 & 0.0205 \\
05 & 0.0004 & 0.0969 & 0.0914 & 0.7706 & 0.0059 & 0.0208 \\
06 & 0.0003 & 0.0054 & 0.0917 & 0.7635 & 0.0053 & 0.0152 \\
07 & 0.0004 & 0.0444 & 0.0916 & 0.7601 & 0.0048 & 0.0104 \\
08 & 0.0169 & 0.0264 & 0.0955 & 0.7643 & 0.0036 & 0.0068 \\
09 & 0.0167 & 0.0662 & 0.0966 & 0.7618 & 0.0031 & 0.0058 \\
10 & 0.0167 & 0.0121 & 0.0975 & 0.7629 & 0.0024 & 0.0053 \\
11 & 0.0168 & 0.0437 & 0.0977 & 0.7649 & 0.0021 & 0.0042 \\
12 & 0.0169 & 0.0208 & 0.0981 & 0.7671 & 0.0018 & 0.0035 \\
13 & 0.0169 & 0.0230 & 0.0985 & 0.7684 & 0.0016 & 0.0028 \\
14 & 0.0169 & 0.0151 & 0.0990 & 0.7686 & 0.0013 & 0.0023 \\
15 & 0.0170 & 0.0338 & 0.0989 & 0.7702 & 0.0011 & 0.0022 \\
16 & 0.0170 & 0.0476 & 0.0990 & 0.7716 & 0.0010 & 0.0020 \\
17 & 0.0171 & 0.0319 & 0.0994 & 0.7731 & 0.0009 & 0.0018 \\
18 & 0.0170 & 0.0001 & 0.0995 & 0.7724 & 0.0008 & 0.0017 \\
19 & 0.0171 & 0.0088 & 0.0994 & 0.7729 & 0.0007 & 0.0016 \\
20 & 0.0171 & 0.0053 & 0.0994 & 0.7736 & 0.0007 & 0.0014 \\
21 & 0.0171 & 0.0825 & 0.0994 & 0.7739 & 0.0006 & 0.0014 \\
22 & 0.0171 & 0.0109 & 0.0995 & 0.7741 & 0.0006 & 0.0013 \\
23 & 0.0171 & 0.0165 & 0.0995 & 0.7741 & 0.0006 & 0.0012 \\
24 & 0.0150 & 0.3716 & 0.0994 & 0.7730 & 0.0005 & 0.0011 \\
25 & 0.0171 & 0.0033 & 0.0995 & 0.7734 & 0.0005 & 0.0010 \\
26 & 0.0170 & 0.0578 & 0.0995 & 0.7732 & 0.0004 & 0.0010 \\
\bottomrule
\end{tabular}
\label{cross-entropy_tab50}
\end{table}

\begin{table}[h!]
\caption{Performance of CE algorithm in the bi-dimensional case for different number of samples.}
\begin{tabular}{ccccc}
\toprule
  samples & levels & CPU time*  & speed-up & function \\
                &          & (seconds)   &                 & evaluation \\
\midrule
   reference  &  --- & 13 101 & ---  & 65 536 \\
              25 & 19    & 106      & 123 &  475 \\
              50 & 26    & 287      &   46 & 1 300\\
              75 & 30    & 497      &   26 & 2 250\\
            100 & 28    & 610      &   21 & 2 800\\
\bottomrule
\end{tabular}
\label{speedup_tab}

\vspace{2mm}
\footnotesize{*\texttt{Dell Inspiron Core i7-3632QM 2.20 GHz RAM 12GB}}
\end{table}

Note that the approximation obtained is very close to the reference value of section~\ref{ref_solution}, obtained after 26 iterative steps (1300 function evaluations), which corresponds to a speed-up of more than 40, when compared with the exhaustive search (see in Table~\ref{speedup_tab} the CPU time spent). The accuracy can still be slightly improved as shown in Table~\ref{cross-entropy_tab75}, which presents the CE results with $N^s = 75$, obtaining $P_{max}^{CE} = 0.0171$. In this case, the price paid for this additional gain of accuracy is a loss of performance, which makes the speed-up fall from more than 40 to 26 (see Table~\ref{speedup_tab}). However, an experiment with only $N^s = 25$ samples is also conducted, obtaining a result with no loss of accuracy and more than doubling the speed-up to an impressive value of 123. An experiment with $N^s = 100$ is also reported in Table~\ref{speedup_tab}, although no figure or table from this simulation is shown in the text. This experiment shows that different sampling strategies can produce good results.

\begin{table}
\caption{Evolution of CE algorithm in the bi-dimensional case using $N^{s} = 75$ samples.}
\begin{tabular}{ccccccc}
\toprule
$\ell$ & $P$ & $K$ & $\mu_{f}$ & $\mu_{\Omega}$ &  $\sigma_{f}$ & $\sigma_{\Omega}$\\
\midrule
01 & 0.0002 & 0.0636 & 0.0868 & 0.7883 & 0.0247 & 0.1244 \\
02 & 0.0003 & 0.1676 & 0.0901 & 0.7712 & 0.0101 & 0.0355 \\
03 & 0.0003 & 0.0224 & 0.0883 & 0.7683 & 0.0074 & 0.0187 \\
04 & 0.0004 & 0.0181 & 0.0936 & 0.7634 & 0.0048 & 0.0119 \\
05 & 0.0004 & 0.0448 & 0.0966 & 0.7597 & 0.0030 & 0.0082 \\
06 & 0.0167 & 0.0358 & 0.0974 & 0.7628 & 0.0021 & 0.0065 \\
07 & 0.0168 & 0.0425 & 0.0978 & 0.7651 & 0.0019 & 0.0052 \\
08 & 0.0169 & 0.0533 & 0.0986 & 0.7679 & 0.0016 & 0.0041 \\
09 & 0.0170 & 0.0473 & 0.0987 & 0.7705 & 0.0013 & 0.0035 \\
10 & 0.0170 & 0.0177 & 0.0988 & 0.7705 & 0.0012 & 0.0029 \\
11 & 0.0170 & 0.0415 & 0.0991 & 0.7706 & 0.0009 & 0.0024 \\
12 & 0.0170 & 0.0519 & 0.0991 & 0.7716 & 0.0008 & 0.0021 \\
13 & 0.0170 & 0.0173 & 0.0992 & 0.7717 & 0.0008 & 0.0021 \\
14 & 0.0170 & 0.0867 & 0.0992 & 0.7727 & 0.0007 & 0.0021 \\
15 & 0.0171 & 0.0349 & 0.0992 & 0.7733 & 0.0006 & 0.0020 \\
16 & 0.0171 & 0.0427 & 0.0992 & 0.7735 & 0.0006 & 0.0018 \\
17 & 0.0171 & 0.0234 & 0.0991 & 0.7736 & 0.0005 & 0.0016 \\
18 & 0.0171 & 0.0077 & 0.0991 & 0.7739 & 0.0005 & 0.0015 \\
19 & 0.0170 & 0.0742 & 0.0991 & 0.7734 & 0.0005 & 0.0013 \\
20 & 0.0171 & 0.0049 & 0.0993 & 0.7736 & 0.0005 & 0.0012 \\
21 & 0.0171 & 0.0085 & 0.0994 & 0.7737 & 0.0004 & 0.0012 \\
22 & 0.0170 & 0.0806 & 0.0995 & 0.7737 & 0.0004 & 0.0013 \\
23 & 0.0171 & 0.0272 & 0.0995 & 0.7742 & 0.0004 & 0.0012 \\
24 & 0.0171 & 0.0370 & 0.0996 & 0.7742 & 0.0004 & 0.0011 \\
25 & 0.0171 & 0.0485 & 0.0996 & 0.7741 & 0.0003 & 0.0011 \\
26 & 0.0171 & 0.0099 & 0.0995 & 0.7745 & 0.0003 & 0.0011 \\
27 & 0.0151 & 0.0857 & 0.0996 & 0.7744 & 0.0003 & 0.0011 \\
28 & 0.0171 & 0.0532 & 0.0996 & 0.7749 & 0.0003 & 0.0010 \\
29 & 0.0171 & 0.0097 & 0.0996 & 0.7742 & 0.0003 & 0.0010 \\
30 & 0.0171 & 0.0512 & 0.0995 & 0.7742 & 0.0003 & 0.0010 \\
\bottomrule
\end{tabular}
\label{cross-entropy_tab75}
\end{table}

\begin{table}
\caption{Evolution of CE algorithm in the bi-dimensional case using $N^{s} = 25$ samples.}
\begin{tabular}{ccccccc}
\toprule
$\ell$ & $P$ & $K$ & $\mu_{f}$ & $\mu_{\Omega}$ &  $\sigma_{f}$ & $\sigma_{\Omega}$\\
\midrule
01 & 0.0003 & 0.1251 & 0.0883 & 0.7870 & 0.0276 & 0.1106 \\
02 & 0.0003 & 0.0095 & 0.0867 & 0.7692 & 0.0131 & 0.0352 \\
03 & 0.0003 & 0.0522 & 0.0889 & 0.7658 & 0.0095 & 0.0181 \\
04 & 0.0003 & 0.0822 & 0.0891 & 0.7608 & 0.0074 & 0.0131 \\
05 & 0.0164 & 0.0347 & 0.0938 & 0.7596 & 0.0052 & 0.0111 \\
06 & 0.0167 & 0.0451 & 0.0960 & 0.7604 & 0.0038 & 0.0077 \\
07 & 0.0167 & 0.0341 & 0.0981 & 0.7620 & 0.0025 & 0.0056 \\
08 & 0.0168 & 0.0415 & 0.0986 & 0.7656 & 0.0022 & 0.0051 \\
09 & 0.0168 & 0.0845 & 0.0985 & 0.7666 & 0.0021 & 0.0049 \\
10 & 0.0169 & 0.0061 & 0.0987 & 0.7674 & 0.0018 & 0.0035 \\
11 & 0.0169 & 0.0007 & 0.0986 & 0.7662 & 0.0015 & 0.0029 \\
12 & 0.0169 & 0.0453 & 0.0988 & 0.7671 & 0.0013 & 0.0023 \\
13 & 0.0170 & 0.0648 & 0.0993 & 0.7684 & 0.0010 & 0.0019 \\
14 & 0.0170 & 0.0569 & 0.0993 & 0.7700 & 0.0009 & 0.0017 \\
15 & 0.0170 & 0.0226 & 0.0992 & 0.7707 & 0.0008 & 0.0014 \\
16 & 0.0170 & 0.0497 & 0.0993 & 0.7710 & 0.0008 & 0.0012 \\
17 & 0.0170 & 0.0716 & 0.0992 & 0.7714 & 0.0007 & 0.0013 \\
18 & 0.0171 & 0.0226 & 0.0996 & 0.7723 & 0.0006 & 0.0012 \\
19 & 0.0171 & 0.0089 & 0.0996 & 0.7730 & 0.0006 & 0.0010 \\
\bottomrule
\end{tabular}
\label{cross-entropy_tab25}
\end{table}

The speed-up shown in Table~\ref{speedup_tab} is defined as the ratio between the calculation time between direct search (reference) and the CE solution. Although this metric provides a good measure of efficiency for large values of CPU times, it is extremely dependent on the machine used. Thus, to provide a machine-independent performance measure, this table also lists the number of evaluations of the objective function, since this is the most expensive operation to be performed by the optimization algorithm. Note that in this new metric, the ratio between the objective function evaluations in the reference and the CE solution produces values very close to the speed-ups obtained previously, confirming the computational efficiency of the proposed approach in a machine-independent fashion.

An animation of the algorithm in action with $N^s = 50$ is available in \cite{video1}, where the reader can see the algorithm stops after just $\ell = 17$ levels (850 function evaluations). The difference concerning the numerical experiment reported in Table~\ref{cross-entropy_tab50} is because the CE method is stochastic so that at each simulation, a different approximation for the global optimum is constructed. Despite these variations, the speed-up and function evaluation values reported in Table~\ref{cross-entropy_tab50} are typical, with some fluctuation around them each new simulation, but with no change in the order of magnitude. 
Animations for the cases where $N^s = 25$ and $N^s = 75$ can be seen in \cite{video3} and \cite{video2}, respectively, where it is possible to note that the accuracy and speed-ups obtained are comparable to those results shown in Tables~\ref{cross-entropy_tab75} and \ref{cross-entropy_tab25}.

\subsection{Noise robustness}

In the practical operation of a vibratory device, noise is inevitable, so considering its effect on the dynamics of energy harvesting systems is good modeling practice \cite{lopes_ccis2019}. Therefore, robustness to noise is a desirable feature in any optimization methodology in problems involving dynamical systems. To test the robustness of the CE solution to noise disturbances, this section slightly modifies the reference solution from section~\ref{ref_solution}. For that, the output voltage signal is corrupted with Gaussian white noise component (zero-mean and standard deviation equal to 5\% of the maximum voltage amplitude). In this way, the objective function of the optimization problem becomes noisy, providing a more stringent test for the optimizer.

\begin{figure}[h]
\centering
\subfigure[]{\includegraphics[scale=0.7]{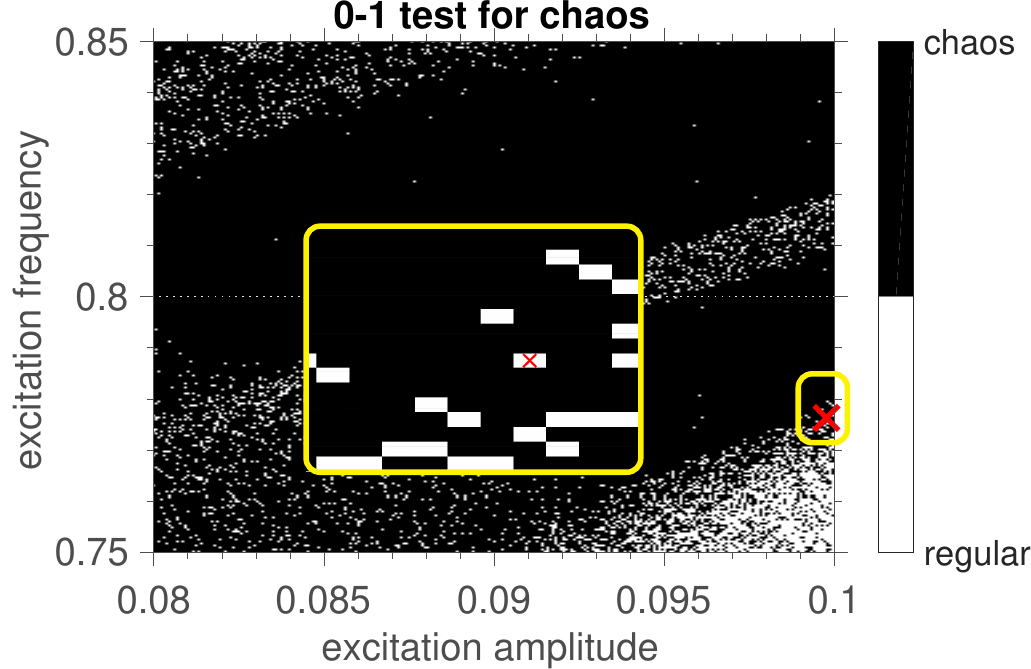}} ~~
\subfigure[]{\includegraphics[scale=0.7]{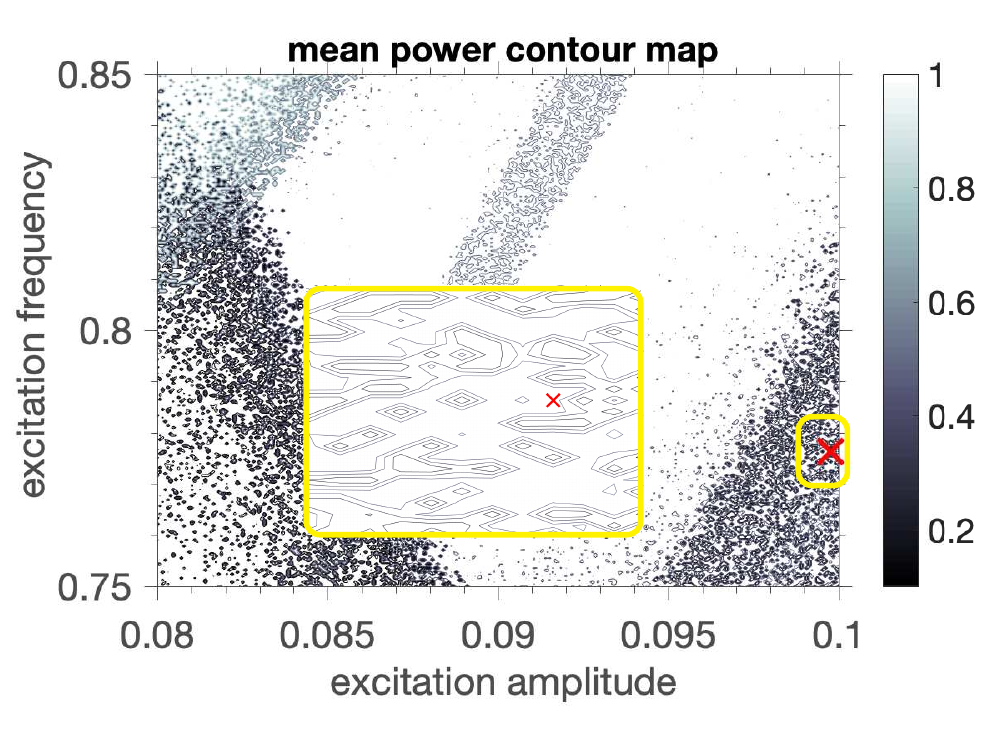}}
\caption{Noisy case contour maps: (a) constraint function defined by the 0-1 test for chaos, and (b) noisy objective function defined by the mean output power.}
\label{ref_solution_noise}
\end{figure}

The contour maps of the constraint and the noisy objective function, with their respective magnifications, both constructed on the same $256 \times 256$ numerical grid from section~\ref{ref_solution} (CPU time 3.8 hours), can be seen in Figure~\ref{ref_solution_noise}. Note that the noise disturbance changes the geometry of the constraint contour map, expanding the regions of chaotic configuration. It also affects the mean power isolines, thus changing a little bit the objective function, so that $P_{max}^{DS} = 0.0173$ at $(f,\Omega) = (0.0998, 0.7763)$.

\begin{figure*}
\centering
\subfigure[ $\ell=1$]{\includegraphics[scale=0.37]{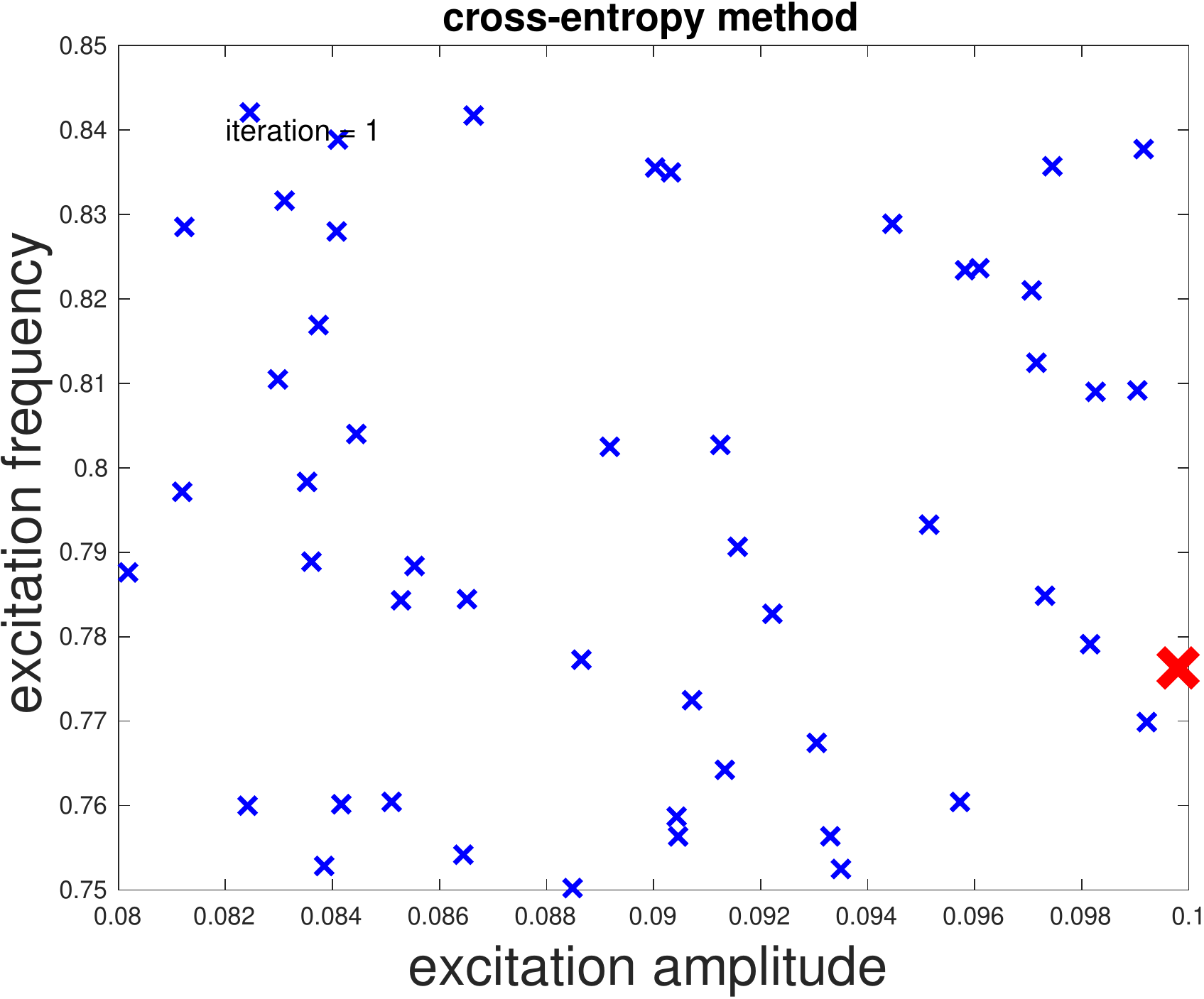}} ~~
\subfigure[ $\ell=5$]{\includegraphics[scale=0.37]{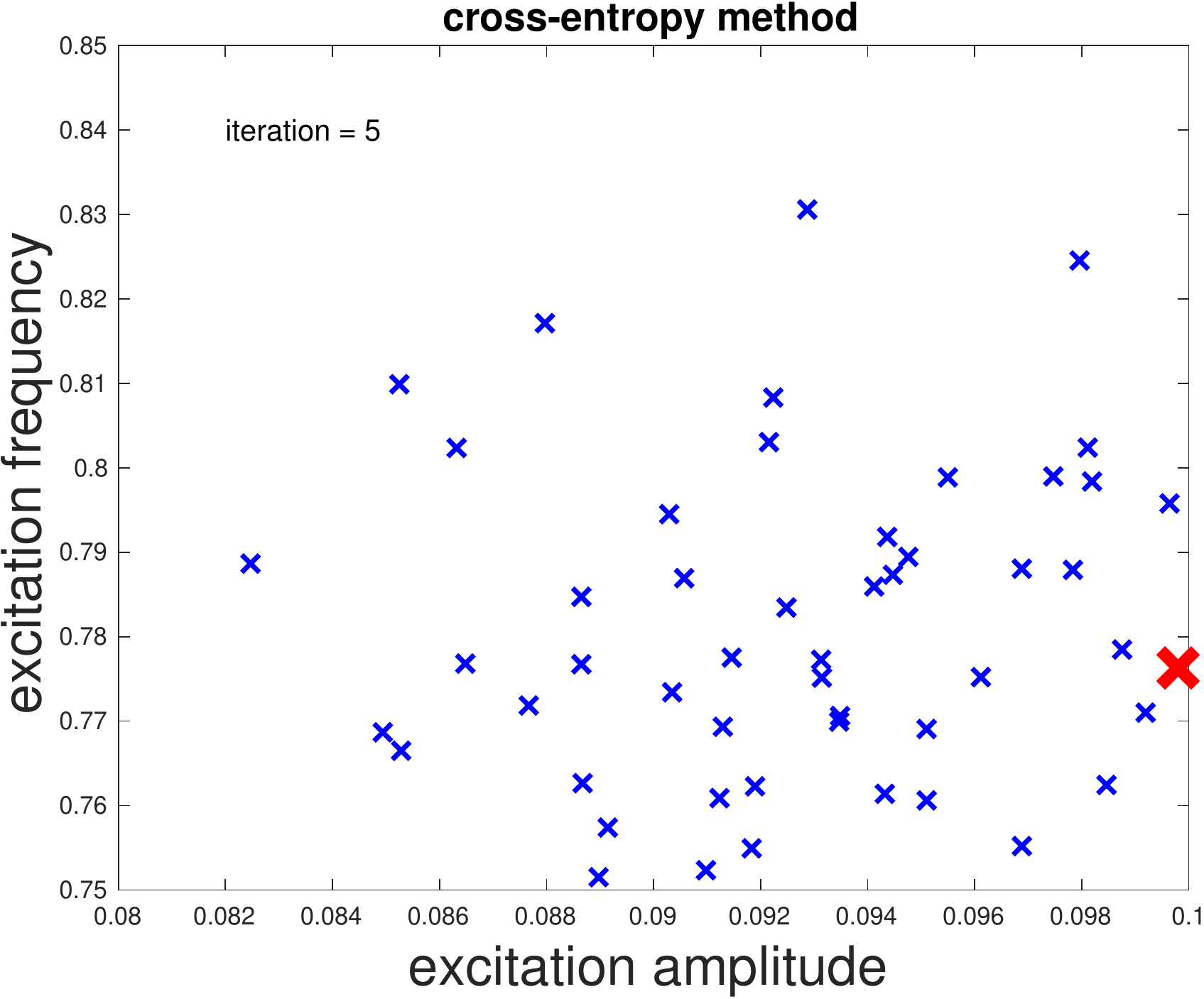}}\\
\subfigure[ $\ell=9$]{\includegraphics[scale=0.37]{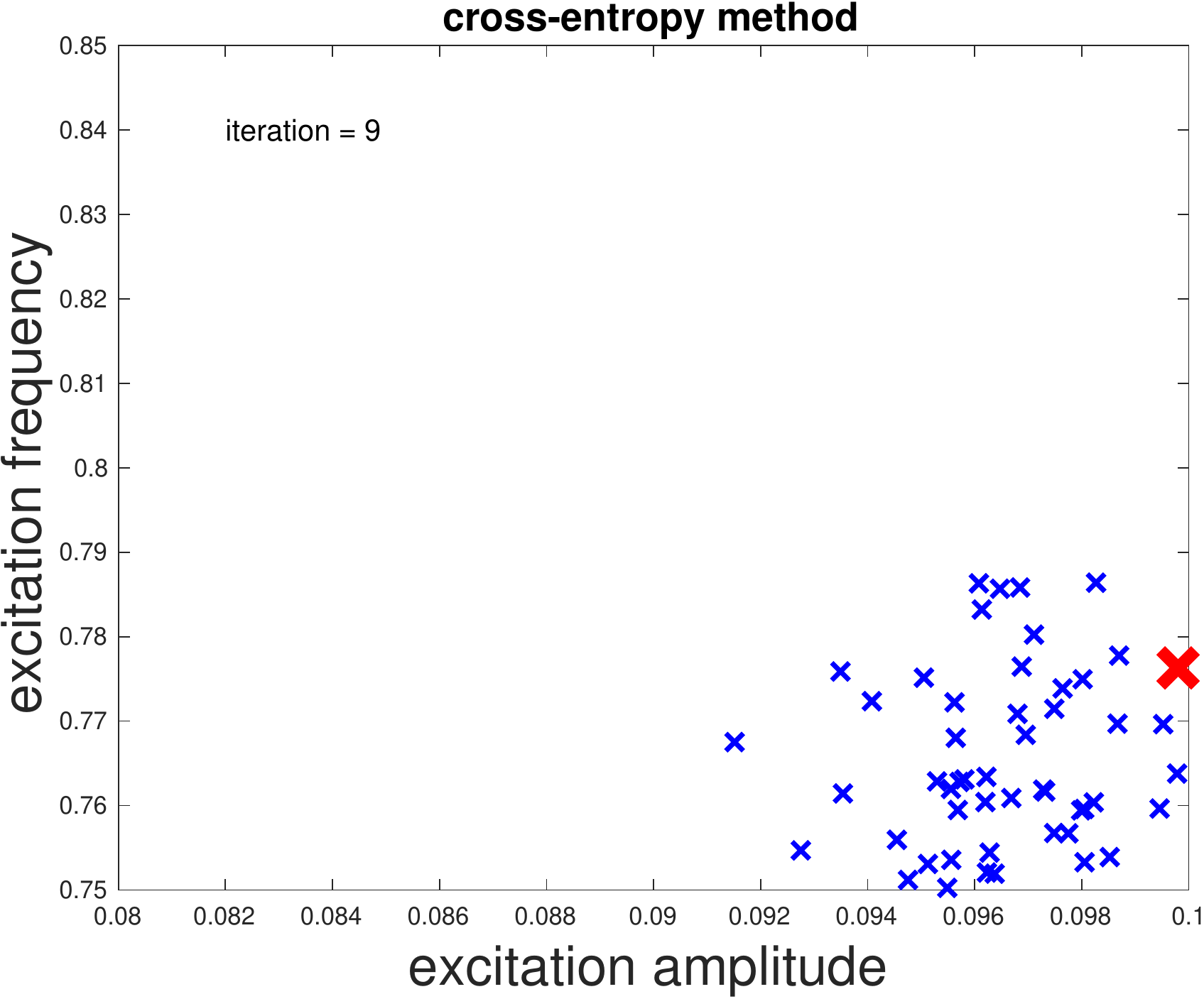}} ~~
\subfigure[$\ell=13$]{\includegraphics[scale=0.37]{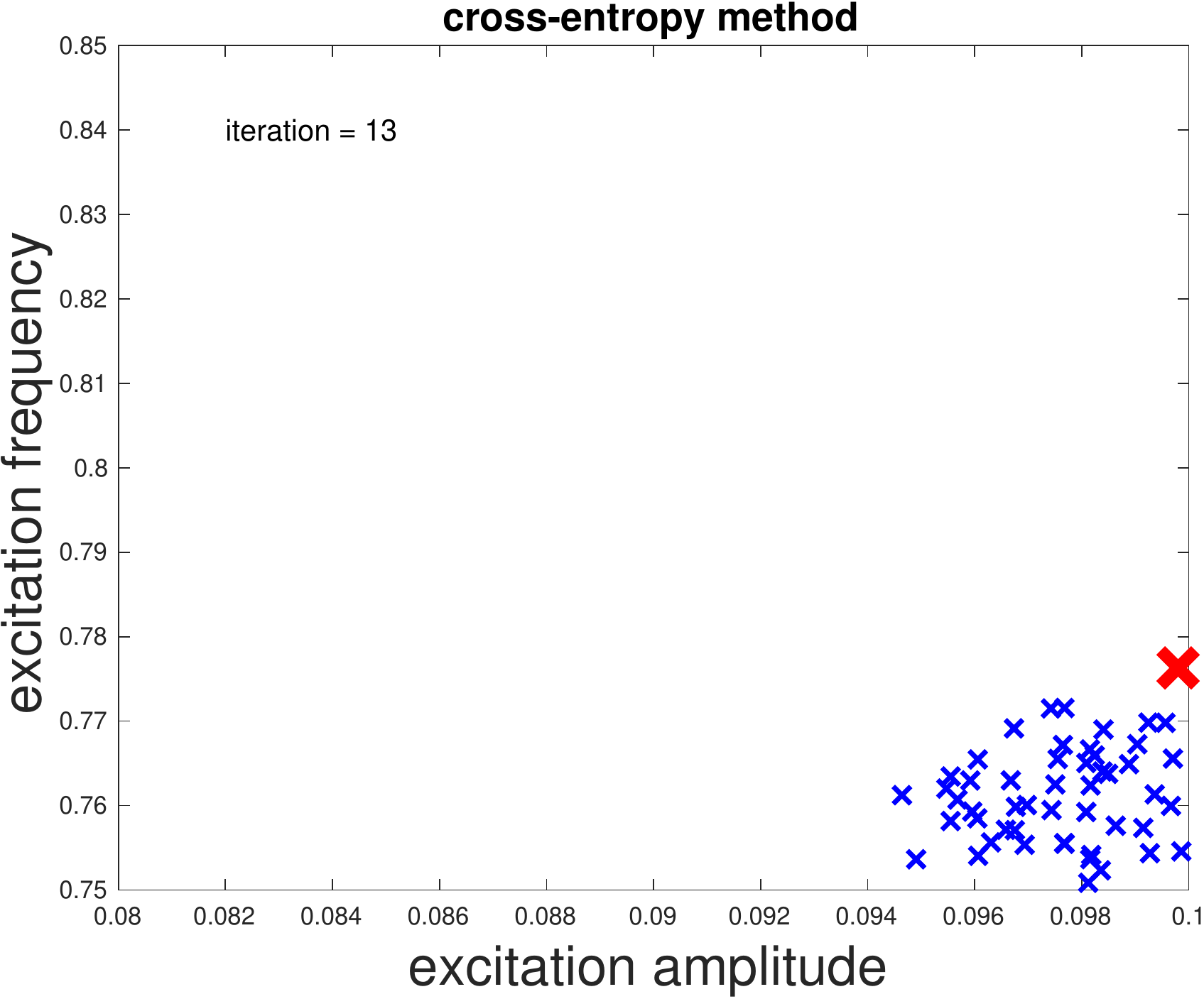}}\\
\subfigure[$\ell=20$]{\includegraphics[scale=0.37]{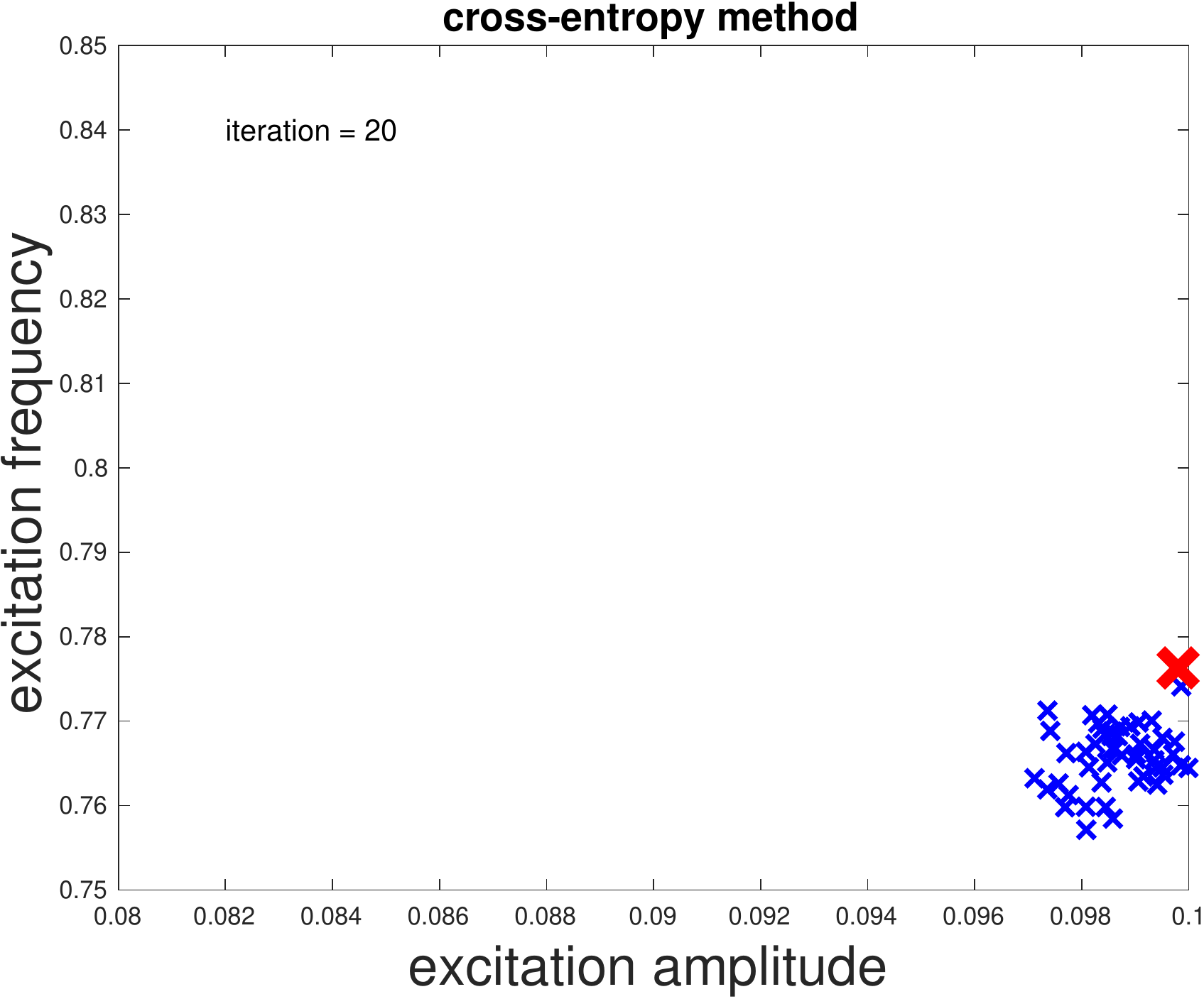}} ~~
\subfigure[$\ell=25$]{\includegraphics[scale=0.37]{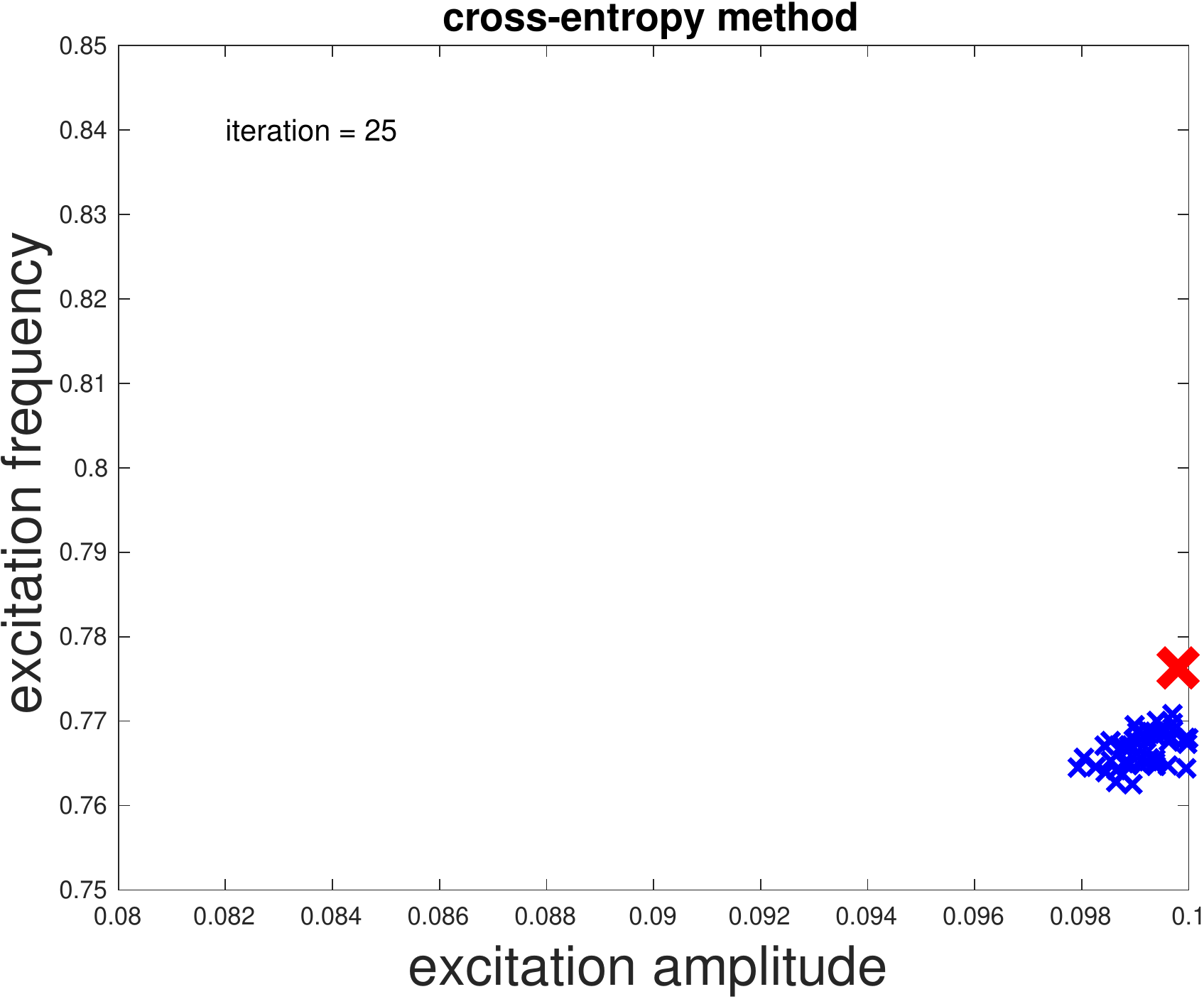}}\\
\caption{Illustration of CE method in the noisy case, using $N^s = 50$ samples, at different levels (iterations) of the algorithm. The noisy version of reference solution is indicated with a red cross.}
\label{cross-entropy_noise}
\end{figure*}

In this noisy experiment, the CE method uses $N^s = 50$, tolerance $\texttt{tol} = 1/512$, and all other parameters set as in the noiseless case. In this way, the CE method finds, after 25 iterations, $P_{max}^{CE} = 0.0170$ at the pair $(f,\Omega) = (0.0991, 0.7675)$, which is a very accurate approximation of the reference solution presented above. An illustration of this iterative process is shown in Figure~\ref{cross-entropy_noise}. Note that here the CE algorithm constructs the approximation for the optimum by a different path than the one shown in Figure~\ref{cross-entropy_fig50}.

Even though the presence of noise complicates the penalized objective function assessment, the example above shows that the CE method may be able to find an accurate approximation for the global optimum. But a caveat needs to be made. Depending on the chosen tolerance value, the CE method may ``froze in a certain region''. For this experiment, for example, a solution with an accuracy defined by $\texttt{tol} = 0.001 <1/512$ cannot be reached in $\ell_{max} = 100$ iterations, as the approximations ``freeze" in a region close to, but not close enough to, the global optimum. In practical terms, this can be a limitation, which depends on the problem at hand.

\subsection{Multidimensional optimization}

In principle, the CE method is extensible to any finite dimension. However, in practice, the CE method suffers from a pathology known as the ``curse of dimensionality'', which is the significant drop of the CE estimator accuracy as the number of design variables increase \cite{Rubinstein2009p547}. Although an improved CE algorithm, that can partially mitigate this problem, exists in the literature  \cite{Rubinstein2009p547}, it is not considered here since the paper aims to explore the simplest, and most intuitive, version of the CE method.

In the context of interest in this paper, optimal design of an energy harvester (and in a broader sense, in dynamic systems), the optimization problems generally have a few design variables, with a dimension of half a dozen being considered high. In this case, although less accurate than in the two-dimensional case, the solution obtained by the CE method can still be useful. 

To illustrate this point, an example with 4 design variables, $\vec{x} = \left( \xi, \chi, \lambda, \kappa \right)$, is considered in this section, where $f=0.115$ and $\Omega = 0.8$ are fixed and the feasible region is defined by $$\mathcal{D} = [0.01, 0.05] \times [0.05, 0.2] \times [0.05, 0.2] \times [0.5, 1.5].$$

A reference solution is constructed via direct search on a $16 \times 16 \times 16 \times 16$ uniform mesh defined in $\mathcal{D}$, where the global maximum $P_{max}^{DS} = 0.1761$ is obtained at $\left( \xi, \chi, \lambda, \kappa \right) = (0.0340, 0.0600, 0.2000, 1.5000)$. The computational budget spent to obtain this solution is similar to that of the direct search in section~\ref{ref_solution}, involving 65 536 evaluations of the objective function. As now the feasible region has a higher dimension than in the two-dimensional example, it is expected that the direct search solution obtained will be less accurate. But this is not problematic for the analyzes presented here, even though it is less precise, this numerical solution is suitable for reference purposes.

\pagebreak
Using the CE method in this case it is possible to speed-up the optimization process and obtain a approximation reasonably close to the global optimum. Table~\ref{speedup_tab2} shows the results obtained for different values of $N^s$, $\texttt{tol} = 0.01$, and considering the other parameters of the algorithm as in section~\ref{ce_num_result}. In all scenarios, the solution obtained by the CE method corresponds to a mean power very close to the one obtained by the direct search in the numerical grid, with speed-up gains up to 30 times approximately. These results indicate that, even though it is susceptible to ``curse of dimensionality'', the CE method can be a very appealing algorithm for problems in moderately higher dimensions, since it can provide accurate approximations and in a very competitive processing time.

\begin{table*}[h!]
\caption{Performance of CE algorithm in the multidimensional case, for different number of samples ($\texttt{tol} = 0.01$).}
\begin{tabular}{cccccccccc}
\toprule
  samples & levels & CPU time*  & speed-up & function    & $P_{max}$  & $\xi$ & $\chi$ & $\lambda$ & $\kappa$ \\
                &          & (seconds)   &                & evaluation &                  &           &            &                   &   \\
\midrule
   reference  &  --- &  12 648 & --- & 65 536 & 0.1761 & 0.0340 & 0.0600 & 0.2000 & 1.5000\\
            100 & 75  &       2074 &    6 &  7 500  & 0.1612 & 0.0237 & 0.1053 & 0.1953 & 1.4923 \\
              50 & 59  &         696 &  18 &  2 950  & 0.1603 & 0.0227 & 0.1099 & 0.1938 & 1.4969 \\
              25 & 83  &         466 &  27 &  2 075  & 0.1619 & 0.0227 & 0.1062 & 0.1955 & 1.4924 \\
\bottomrule
\end{tabular}
\label{speedup_tab2}

\vspace{2mm}
\footnotesize{*\texttt{Dell Inspiron Core i7-3632QM 2.20 GHz RAM 12GB}}
\end{table*}

\subsection{Remarks on CE solution}

The above results allow one to conclude that the optimization approach based on the CE method is robust (able to find the global optimal) and efficient (computationally feasible) to address this nonlinear non-convex optimization problem, which has non-trivial numerical solution since the existence of jump-like discontinuities in the constraint prevents gradient-based algorithms from being used. 

Last but not least, it is worth mentioning that the CE method has a great advantage in terms of simplicity when compared to most of the meta-heuristics used in non-differentiable optimization since the algorithm involves only four control parameters, all of which have a very intuitive meaning, which is a considerable advantage compared to genetic algorithms for example.


\section{Conclusions}
\label{concl_sect}

This work presented the formulation of a nonlinear non-convex optimization problem that seeks to maximize the efficiency of a bistable energy harvesting system driven by a sinusoidal excitation and constrained to operate in non-chaotic dynamic regimes only. Since the problem has jump-type discontinuities, which prevents the use of gradient-based methods, a stochastic strategy of optimization, based on the cross-entropy method, was proposed to construct a numerical approximation for the optimal solution. 

Tests to verify the efficiency and accuracy of this cross-entropy approach as well as its robustness to noise and feasibility of use in larger dimensions, were conducted. They showed that the proposed strategy of optimization is quite robust and effective, constituting a very appealing tool to deal with the typical (non-convex) problems related to the optimization of energy harvesting systems, even in the presence of noise or in a problem with moderately high dimension.

The impressive results reported here can still be improved in terms of performance, through the use of parallelization strategies. In particular, it would be interesting to test the cloud computing parallelization strategy proposed by \cite{cunhajr2014p1355}. Besides, the optimization framework is extremely versatile, allowing the extension to problems involving robust optimization such as those reported by \cite{cunhajr2015p849,cunhajr2019meccanica}, in which the global optimum is not sought in the strict sense, but in a sense in which the system performance is maximized in average terms, respecting probabilistic constraints.

\section*{Acknowledgements}

Including noise into the objective function is an idea given to the author for the first time in a conference at Morocco by Prof. Luca Gammaitoni (University of Perugia), and more recently by one of the anonymous reviewers. The noisily test case proved to be very useful to illustrate the robustness of the framework presented here. The author is very grateful to both of them for this suggestion. He is also indebted to Dr. Welington de Oliveira (MINES ParisTech), for the fruitful discussions about the mathematical technicalities related to the optimization problem addressed in this work.

\section*{Funding}

This research received financial support from the Brazilian agencies Coordena\c{c}\~{a}o de Aperfei\c{c}oamento de Pessoal de N\'{\i}vel Superior - Brasil (CAPES) - Finance Code 001, and the Carlos Chagas Filho Research Foundation of Rio de Janeiro State (FAPERJ) under the following grants: 211.304/2015, 210.021/2018, 210.167/2019, and 211.037/2019. 

\section*{Compliance with ethical standards}

\section*{Conflict of Interest }
The author declares that he has no conflict of interest.

\pagebreak
\section*{Code availability}

To facilitate the reproduction of this paper results, as well as to popularize the CE method use by the community of nonlinear dynamical systems, the code used in the simulations is available in the following repository:\\
\textcolor{magenta}{\url{https://americocunhajr.github.io/HarvesterOpt}}


\bibliographystyle{spbasic}      

\bibliography{references}

\begin{thebibliography}{68}
\providecommand{\natexlab}[1]{#1}
\providecommand{\url}[1]{{#1}}
\providecommand{\urlprefix}{URL }
\expandafter\ifx\csname urlstyle\endcsname\relax
  \providecommand{\doi}[1]{DOI~\discretionary{}{}{}#1}\else
  \providecommand{\doi}{DOI~\discretionary{}{}{}\begingroup
  \urlstyle{rm}\Url}\fi
\providecommand{\eprint}[2][]{\url{#2}}

\bibitem[{Abdelkefi et~al.(2012)Abdelkefi, Nayfeh, and
  Hajj}]{Abdelkefi2012p531}
Abdelkefi A, Nayfeh AH, Hajj MR (2012) Enhancement of power harvesting from
  piezoaeroelastic systems. Nonlinear Dynamics 68:531--541,
  \doi{https://doi.org/10.1007/s11071-011-0234-9}

\bibitem[{Barbosa et~al.(2015)Barbosa, {De Paula}, Savi, and
  Inman}]{Savi2015p2787}
Barbosa WOV, {De Paula} AS, Savi MA, Inman DJ (2015) Chaos control applied to
  piezoelectric vibration-based energy harvesting systems. The European
  Physical Journal Special Topics 224:2787--2801,
  \doi{https://doi.org/10.1140/epjst/e2015-02589-1}

\bibitem[{Benacchio et~al.(2016)Benacchio, Malher, Boisson, and
  Touz{\'e}}]{Benacchio2016p893}
Benacchio S, Malher A, Boisson J, Touz{\'e} C (2016) Design of a magnetic
  vibration absorber with tunable stiffnesses. Nonlinear Dynamics 85:893--911,
  \doi{https://doi.org/10.1007/s11071-016-2731-3}

\bibitem[{Bernardini and Litak(2016)}]{litak2016p1433}
Bernardini D, Litak G (2016) An overview of 0–1 test for chaos. Journal of
  the Brazilian Society of Mechanical Sciences and Engineering 38:1433–1450,
  \doi{https://doi.org/10.1007/s40430-015-0453-y}

\bibitem[{Bhatti et~al.(2016)Bhatti, Alizai, Syed, and L}]{Bhatti2016p24}
Bhatti NA, Alizai MH, Syed AA, L M (2016) Energy harvesting and wireless
  transfer in sensor network applications: Concepts and experiences. ACM
  Transactions on Sensor Networks 12:24:1--24:40,
  \doi{https://doi.org/10.1145/2915918}

\bibitem[{Bonnans et~al.(2009)Bonnans, Gilbert, Lemarechal, and
  Sagastiz\'{a}bal}]{bonnans2009}
Bonnans J, Gilbert JC, Lemarechal C, Sagastiz\'{a}bal CA (2009) Numerical
  Optimization: Theoretical and Practical Aspects, 2nd edn. Springer

\bibitem[{Boyd and Vandenberghe(2004)}]{boyd2004}
Boyd S, Vandenberghe L (2004) Convex Optimization. Cambridge University Press

\bibitem[{Catacuzzeno et~al.(2019)Catacuzzeno, Orfei, Michele, Sforna,
  Franciolini, and Gammaitoni}]{Catacuzzeno2019p823}
Catacuzzeno L, Orfei F, Michele AD, Sforna L, Franciolini F, Gammaitoni L
  (2019) Energy harvesting from a bio cell. Nano Energy 56:823--827

\bibitem[{Cottone et~al.(2009)Cottone, Vocca, and
  Gammaitoni}]{cottone2009p080601}
Cottone F, Vocca H, Gammaitoni L (2009) Nonlinear energy harvesting. Phys Rev
  Lett 102:080601, \doi{https://doi.org/10.1103/PhysRevLett.102.080601}

\bibitem[{{Cunha~Jr}(2020{\natexlab{a}})}]{video3}
{Cunha~Jr} A (2020{\natexlab{a}}) Cross-entropy optimization of bistable energy
  harvesting system (25 samples). \url{https://youtu.be/0EvzdVXlPqA}, (Accessed
  28 March 2020)

\bibitem[{{Cunha~Jr}(2020{\natexlab{b}})}]{video1}
{Cunha~Jr} A (2020{\natexlab{b}}) Cross-entropy optimization of bistable energy
  harvesting system (50 samples). \url{https://youtu.be/-JB3eniIdDY}, (Accessed
  28 March 2020)

\bibitem[{{Cunha~Jr}(2020{\natexlab{c}})}]{video2}
{Cunha~Jr} A (2020{\natexlab{c}}) Cross-entropy optimization of bistable energy
  harvesting system (75 samples). \url{https://youtu.be/uIZM4SjCbrw}, (Accessed
  28 March 2020)

\bibitem[{{Cunha~Jr} et~al.(2014){Cunha~Jr}, Nasser, Sampaio, Lopes, and
  Breitman}]{cunhajr2014p1355}
{Cunha~Jr} A, Nasser R, Sampaio R, Lopes H, Breitman K (2014) Uncertainty
  quantification through {M}onte {C}arlo method in a cloud computing setting.
  Computer Physics Communications 185:1355--1363,
  \doi{https://doi.org/10.1016/j.cpc.2014.01.006}

\bibitem[{{Cunha~Jr} et~al.(2015){Cunha~Jr}, Soize, and
  Sampaio}]{cunhajr2015p849}
{Cunha~Jr} A, Soize C, Sampaio R (2015) Computational modeling of the nonlinear
  stochastic dynamics of horizontal drillstrings. Computational Mechanics
  56:849--878, \doi{https://doi.org/10.1007/s00466-015-1206-6}

\bibitem[{Dantas(2019)}]{Dantas2019}
Dantas E (2019) A cross-entropy strategy for parameters identification
  problems. Monograph, Universidade do Estado do Rio de Janeiro,
  https://dx.doi.org/10.13140/RG.2.2.18045.51688

\bibitem[{Dantas et~al.(2019{\natexlab{a}})Dantas, {Cunha~Jr}, and
  Silva}]{cunhajr_icedyn2019}
Dantas E, {Cunha~Jr} A, Silva TAN (2019{\natexlab{a}}) A numerical procedure
  based on cross-entropy method for stiffness identification. In: 5th
  International Conference on Structural Engineering Dynamics (ICEDyn~2019),
  Viana do Castelo, Portugal

\bibitem[{Dantas et~al.(2019{\natexlab{b}})Dantas, {Cunha~Jr}, Soeiro, Cayres,
  and Weber}]{cunhajr_cobem2019_2}
Dantas E, {Cunha~Jr} A, Soeiro FJCP, Cayres BC, Weber HI (2019{\natexlab{b}})
  An inverse problem via cross-entropy method for calibration of a drill string
  torsional dynamic model. In: 25th ABCM International Congress of Mechanical
  Engineering (COBEM~2019), Uberl\^{a}ndia, Brazil,
  \doi{http://dx.doi.org/10.26678/abcm.cobem2019.cob2019-2216}

\bibitem[{Daqaq et~al.(2020)Daqaq, Crespo, and Ha}]{Daqaq2020p1525}
Daqaq MF, Crespo RS, Ha S (2020) On the efficacy of charging a battery using a
  chaotic energy harvester. Nonlinear Dynamics 99:1525--1537,
  \doi{https://doi.org/10.1007/s11071-019-05372-0}

\bibitem[{{De Boer} et~al.(2005){De Boer}, Kroese, Mannor, and
  Rubinstein}]{DeBoer2005p19}
{De Boer} P, Kroese DP, Mannor S, Rubinstein RY (2005) A tutorial on the
  cross-entropy method. Annals of Operations Research 134:19--67,
  \doi{https://doi.org/10.1007/s10479-005-5724-z}

\bibitem[{{de la Roca} et~al.(2019){de la Roca}, Peterson, Pereira, and
  {Cunha~Jr}}]{cunhajr_cobem2019_1}
{de la Roca} L, Peterson JVLL, Pereira MC, {Cunha~Jr} A (2019) Control of chaos
  via {OGY} method on a bistable energy harvester. In: 25th ABCM International
  Congress of Mechanical Engineering (COBEM~2019), Uberl\^{a}ndia, Brazil,
  \doi{http://dx.doi.org/10.26678/abcm.cobem2019.cob2019-1970}

\bibitem[{Dekemele et~al.(2019)Dekemele, Van~Torre, and
  Loccufier}]{Dekemele2019p1831}
Dekemele K, Van~Torre P, Loccufier M (2019) Performance and tuning of a chaotic
  bi-stable nes to mitigate transient vibrations. Nonlinear Dynamics
  98:1831--1851, \doi{https://doi.org/10.1007/s11071-019-05291-0}

\bibitem[{Erturk et~al.(2009)Erturk, Hoffmann, and Inman}]{erturk2009p254102}
Erturk A, Hoffmann J, Inman DJ (2009) A piezomagnetoelastic structure for
  broadband vibration energy harvesting. Applied Physics Letters 94:254102,
  \doi{https://doi.org/10.1063/1.3159815}

\bibitem[{Gallo et~al.(2012)Gallo, Tofoli, Rade, and {Steffen,
  Jr}}]{Rade2012p1650}
Gallo CA, Tofoli FL, Rade DA, {Steffen, Jr} V (2012) Piezoelectric actuators
  applied to neutralize mechanical vibrations. Journal of Vibration and Control
  18:1650--1660, \doi{https://doi.org/10.1177/1077546311422549}

\bibitem[{Gammaitoni(2012)}]{Gammaitoni2012p627}
Gammaitoni L (2012) There’s plenty of energy at the bottom (micro and nano
  scale nonlinear noise harvesting). Contemporary Physics 53:119--135,
  \doi{https://doi.org/10.1080/00107514.2011.647793}

\bibitem[{Ghidey(2015)}]{Ghidey2015}
Ghidey H (2015) Reliability-based design optimization with cross-entropy
  method. Master’s thesis, Norwegian University of Science and Technology,
  Trondheim

\bibitem[{Ghouli et~al.(2017)Ghouli, Hamdi, and Belhaq}]{Ghouli2017p1625}
Ghouli Z, Hamdi M, Belhaq M (2017) Energy harvesting from quasi-periodic
  vibrations using electromagnetic coupling with delay. Nonlinear Dynamics
  89:1625--1636, \doi{https://doi.org/10.1007/s11071-017-3539-5}

\bibitem[{Godoy and Trindade(2012)}]{trindade2012p552}
Godoy TC, Trindade MA (2012) Effect of parametric uncertainties on the
  performance of a piezoelectric energy harvesting device. Journal of the
  Brazilian Society of Mechanical Sciences and Engineering 34:552--560,
  \doi{https://dx.doi.org/10.1590/S1678-58782012000600003}

\bibitem[{Gottwald and Melbourne(2004)}]{gottwald2004p603}
Gottwald GA, Melbourne I (2004) A new test for chaos in deterministic systems.
  Proceedings of the Royal Society of London Series A 460:603--611,
  \doi{https://doi.org/10.1098/rspa.2003.1183}

\bibitem[{Gottwald and Melbourne(2009{\natexlab{a}})}]{gottwald2009p129}
Gottwald GA, Melbourne I (2009{\natexlab{a}}) On the implementation of the 0-1
  test for chaos. SIAM Journal on Applied Dynamical Systems 8:129--145,
  \doi{https://doi.org/10.1137/080718851}

\bibitem[{Gottwald and Melbourne(2009{\natexlab{b}})}]{gottwald2009p1367}
Gottwald GA, Melbourne I (2009{\natexlab{b}}) On the validity of the 0-1 test
  for chaos. Nonlinearity 22:1367--1382,
  \doi{https://doi.org/10.1088/0951-7715/22/6/006}

\bibitem[{Gottwald and Melbourne(2016)}]{gottwald2016}
Gottwald GA, Melbourne I (2016) The 0-1 Test for Chaos: A review, vol 915,
  Springer. \doi{https://doi.org/10.1007/978-3-662-48410-4}

\bibitem[{Harne(2012)}]{harne2012p162}
Harne RL (2012) Theoretical investigations of energy harvesting efficiency from
  structural vibrations using piezoelectric and electromagnetic oscillators.
  The Journal of the Acoustical Society of America 132:162--172,
  \doi{https://doi.org/10.1121/1.4725765}

\bibitem[{Ibrahim et~al.(2016)Ibrahim, Towfighian, Younis, and
  Su}]{Ibrahim2016}
Ibrahim A, Towfighian S, Younis M, Su Q (2016) {Magnetoelastic beam with
  extended polymer for low frequency vibration energy harvesting}. In:
  Meyendorf NG, Matikas TE, Peters KJ (eds) Smart Materials and Nondestructive
  Evaluation for Energy Systems 2016, International Society for Optics and
  Photonics, SPIE, vol 9806, pp 71 -- 85,
  \doi{https://doi.org/10.1117/12.2219276}

\bibitem[{Issa et~al.(2018)Issa, {Cunha~Jr}, Soeiro, and
  Pereira}]{cunhajr_cnmac2018}
Issa MVS, {Cunha~Jr} A, Soeiro FJCP, Pereira A (2018) Structural optimization
  using the cross-entropy method. In: {XXXVIII} Congresso Nacional de
  Matem\'{a}tica Aplicada e Computacional (CNMAC~2018), Campinas, Brazil,
  \doi{https://dx.doi.org/10.5540/03.2018.006.02.0443}

\bibitem[{Kroese et~al.(2011)Kroese, Taimre, and Botev}]{kroese2011}
Kroese DP, Taimre T, Botev ZI (2011) Handbook of Monte Carlo Methods. Wiley

\bibitem[{Kroese et~al.(2013)Kroese, Rubinstein, Cohen, Porotsky, and
  Taimre}]{Kroese2013}
Kroese DP, Rubinstein RY, Cohen I, Porotsky S, Taimre T (2013) Cross-Entropy
  Method, Springer, pp 326--333.
  \doi{https://doi.org/10.1007/978-1-4419-1153-7_131}

\bibitem[{Leadenham and Erturk(2020)}]{Leadenham2020p625}
Leadenham S, Erturk A (2020) Mechanically and electrically nonlinear non-ideal
  piezoelectric energy harvesting framework with experimental validations.
  Nonlinear Dynamics 99:625--641,
  \doi{https://doi.org/10.1007/s11071-019-05091-6}

\bibitem[{Liu et~al.(2019)Liu, Xu, Chen, Liu, Li, Liu, and
  Chen}]{Liu2019p109581}
Liu W, Xu X, Chen F, Liu Y, Li S, Liu L, Chen Y (2019) A review of research on
  the closed thermodynamic cycles of ocean thermal energy conversion. Renewable
  and Sustainable Energy Reviews p 109581,
  \doi{https://doi.org/10.1016/j.rser.2019.109581}

\bibitem[{Lopes et~al.(2017)Lopes, Peterson, and
  {Cunha~Jr}}]{lopes_cnmac2017_2}
Lopes VG, Peterson JVLL, {Cunha~Jr} A (2017) Numerical study of parameters
  influence over the dynamics of a piezo-magneto-elastic energy harvesting
  device. In: {XXXVII} {C}ongresso {N}acional de {M}atem\'{a}tica {A}plicada e
  {C}omputacional (CNMAC 2017), S\~{a}o Jos\'{e} dos Campos, Brazil,
  \doi{http://dx.doi.org/10.5540/03.2018.006.01.0407}

\bibitem[{Lopes et~al.(2019{\natexlab{a}})Lopes, Peterson, and
  {Cunha~Jr}}]{cunhajr_belhaq2019}
Lopes VG, Peterson JVLL, {Cunha~Jr} A (2019{\natexlab{a}}) {N}onlinear
  {C}haracterization of a {B}istable {E}nergy {H}arvester {D}ynamical {S}ystem.
  In: Belhaq M (ed) Topics in Nonlinear Mechanics and Physics: Selected Papers
  from CSNDD 2018 (Springer Proceedings in Physics), Springer, Singapore, pp
  71--88, \doi{https://dx.doi.org/10.1007/978-981-13-9463-8_3}

\bibitem[{Lopes et~al.(2019{\natexlab{b}})Lopes, Peterson, and
  {Cunha~Jr}}]{lopes_ccis2019}
Lopes VG, Peterson JVLL, {Cunha~Jr} A (2019{\natexlab{b}}) The nonlinear
  dynamics of a bistable energy harvesting system with colored noise
  disturbances. Journal of Computational Interdisciplinary Sciences 10:125

\bibitem[{L\'opez-Su\'arez et~al.(2011)L\'opez-Su\'arez, Rurali, Gammaitoni,
  and Abadal}]{Gammaitoni2011p161401}
L\'opez-Su\'arez M, Rurali R, Gammaitoni L, Abadal G (2011) Nanostructured
  graphene for energy harvesting. Physical Review B 84:161401,
  \doi{https://doi.org/10.1103/PhysRevB.84.161401}

\bibitem[{Mangla et~al.(2020)Mangla, Ahmad, and Uddin}]{Mangla2020}
Mangla C, Ahmad M, Uddin M (2020) Optimization of complex nonlinear systems
  using genetic algorithm. International Journal of Information Technology
  \doi{10.1007/s41870-020-00421-z}

\bibitem[{Nabavi and Zhang(2016)}]{Nabavi2016}
Nabavi S, Zhang L (2016) {MEMS} piezoelectric energy harvester design and
  optimization based on {G}enetic {A}lgorithm. In: 2016 IEEE International
  Ultrasonics Symposium (IUS), pp 1--4,
  \doi{https://doi.org/10.1109/ULTSYM.2016.7728786}

\bibitem[{Nocedal and Wright(2006)}]{nocedal2006}
Nocedal J, Wright S (2006) Numerical Optimization, 2nd edn. Springer

\bibitem[{Peterson et~al.(2016)Peterson, Lopes, and
  {Cunha~Jr}}]{peterson_cnmac2016}
Peterson JVLL, Lopes VG, {Cunha~Jr} A (2016) Maximization of the electrical
  power generated by a piezo-magneto-elastic energy harvesting device. In:
  {XXXVI} {C}ongresso {N}acional de {M}atem\'{a}tica {A}plicada e
  {C}omputacional (CNMAC~2016), Gramado, Brazil,
  \doi{http://dx.doi.org/10.5540/03.2017.005.01.0200}

\bibitem[{Pfenniger et~al.(2014)Pfenniger, Stahel, Koch, Obrist, and
  Vogel}]{pfenniger2014p3325}
Pfenniger A, Stahel A, Koch VM, Obrist D, Vogel R (2014) Energy harvesting
  through arterial wall deformation: A {FEM} approach to fluid–structure
  interactions and magneto-hydrodynamics. Applied Mathematical Modelling
  38:3325--3338, \doi{https://doi.org/10.1016/j.apm.2013.11.051}

\bibitem[{Priya and Inman(2009)}]{priya2009}
Priya S, Inman DJ (2009) Energy Harvesting Technologies. Springer

\bibitem[{Quaranta et~al.(2020)Quaranta, Lacarbonara, and Masri}]{Quaranta2020}
Quaranta G, Lacarbonara W, Masri SF (2020) A review on computational
  intelligence for identification of nonlinear dynamical systems. Nonlinear
  Dynamics 99:1709--1761, \doi{10.1007/s11071-019-05430-7}

\bibitem[{Ramlan et~al.(2010)Ramlan, Brennan, Mace, and
  Kovacic}]{Ramlan2010p545}
Ramlan R, Brennan MJ, Mace BR, Kovacic I (2010) Potential benefits of a
  non-linear stiffness in an energy harvesting device. Nonlinear Dynamics
  59:545--558, \doi{https://doi.org/10.1007/s11071-009-9561-5}

\bibitem[{Rechenbach et~al.(2016)Rechenbach, Willatzen, and
  Lassen}]{rechenbach2016p1232}
Rechenbach B, Willatzen M, Lassen B (2016) Theoretical study of the
  electromechanical efficiency of a loaded tubular dielectric elastomer
  actuator. Applied Mathematical Modelling 40:1232--1246,
  \doi{https://doi.org/10.1016/j.apm.2015.06.029}

\bibitem[{Rocha et~al.(2019)Rocha, Balthazar, Tusset, de~Souza, Janzen, and
  Arbex}]{Rocha2019p3423}
Rocha RT, Balthazar JM, Tusset AM, de~Souza SLT, Janzen FC, Arbex HC (2019) On
  a non-ideal magnetic levitation system: nonlinear dynamical behavior and
  energy harvesting analyses. Nonlinear Dynamics 95:3423--3438,
  \doi{https://doi.org/10.1007/s11071-019-04765-5}

\bibitem[{Rubinstein(1997)}]{Rubinstein1997p89}
Rubinstein RY (1997) Optimization of computer simulation models with rare
  events. European Journal of Operations Research 99:89--112,
  \doi{https://doi.org/10.1016/S0377-2217(96)00385-2}

\bibitem[{Rubinstein(1999)}]{Rubinstein1999p127}
Rubinstein RY (1999) The cross-entropy method for combinatorial and continuous
  optimization. Methodology and Computing in Applied Probability 2:127--190,
  \doi{https://doi.org/10.1023/A:1010091220143}

\bibitem[{Rubinstein and Glynn(2009)}]{Rubinstein2009p547}
Rubinstein RY, Glynn PW (2009) How to deal with the curse of dimensionality of
  likelihood ratios in {M}onte {C}arlo simulation. Stochastic Models
  25(4):547--568, \doi{10.1080/15326340903291248}

\bibitem[{Rubinstein and Kroese(2004)}]{Rubinstein2004}
Rubinstein RY, Kroese DP (2004) The Cross-Entropy Method: A Unified Approach to
  Combinatorial Optimization, Monte-Carlo Simulation and Machine Learning.
  Information Science and Statistics, Springer-Verlag

\bibitem[{Rubinstein and Kroese(2016)}]{Rubinstein2016}
Rubinstein RY, Kroese DP (2016) Simulation and the Monte Carlo Method, 3rd edn.
  Wiley Series in Probability and Statistics, Wiley

\bibitem[{Selvan and Ali(2016)}]{Selvan2016p1035}
Selvan KV, Ali MSM (2016) Micro-scale energy harvesting devices: {R}eview of
  methodological performances in the last decade. Renewable and Sustainable
  Energy Reviews 54:1035--1047,
  \doi{https://doi.org/10.1016/j.rser.2015.10.046}

\bibitem[{Spies et~al.(2015)Spies, Pollak, and Mateu}]{spies2015}
Spies P, Pollak M, Mateu L (2015) Handbook of Energy Harvesting Power Supplies
  and Applications. Pan Stanford

\bibitem[{Sun et~al.(2020)Sun, {Jianbin}, {Karimi}, and {Fu}}]{Sun2020}
Sun K, {Jianbin} Q, {Karimi} HR, {Fu} Y (2020) Event-triggered robust fuzzy
  adaptive finite-time control of nonlinear systems with prescribed
  performance. IEEE Transactions on Fuzzy Systems pp 1--1

\bibitem[{Trindade(2016)}]{Trindade2016}
Trindade MA (2016) Passive and Active Structural Vibration Control, Springer
  International Publishing, pp 65--92.
  \doi{https://doi.org/10.1007/978-3-319-29982-2_4}

\bibitem[{Vocca et~al.(2012)Vocca, Neri, Travasso, and
  Gammaitoni}]{Vocca2012p771}
Vocca H, Neri I, Travasso F, Gammaitoni L (2012) Kinetic energy harvesting with
  bistable oscillators. Applied Energy 97:771--776,
  \doi{https://doi.org/10.1016/j.apenergy.2011.12.087}

\bibitem[{Wang(2012)}]{Wang2012}
Wang B (2012) Parameter estimation for {ODEs} using a cross-entropy approach.
  Master’s thesis, University of Toronto, Toronto

\bibitem[{Wolszczak et~al.(2019)Wolszczak, Lonkwic, {Cunha~Jr}, Litak, and
  Molski}]{cunhajr2019meccanica}
Wolszczak P, Lonkwic P, {Cunha~Jr} A, Litak G, Molski S (2019) Robust
  optimization and uncertainty quantification in the nonlinear mechanics of an
  elevator brake system. Meccanica 54:1057--1069,
  \doi{https://doi.org/10.1007/s11012-019-00992-7}

\bibitem[{Yang and Cao(2019)}]{Yang2019p1511}
Yang T, Cao Q (2019) Time delay improves beneficial performance of a novel
  hybrid energy harvester. Nonlinear Dynamics 96:1511--1530,
  \doi{https://doi.org/10.1007/s11071-019-04868-z}

\bibitem[{Ying et~al.(2015)Ying, Yuan, and Hu}]{ying2015v7}
Ying Q, Yuan W, Hu N (2015) Improving the efficiency of harvesting electricity
  from living trees. Journal of Renewable and Sustainable Energy 7,
  \doi{https://doi.org/10.1063/1.4935577}

\bibitem[{Zheng et~al.(2009)Zheng, Chang, and Gea}]{Zheng2009p17}
Zheng B, Chang CJ, Gea HC (2009) Topology optimization of energy harvesting
  devices using piezoelectric materials. Structural and Multidisciplinary
  Optimization 38:17--23, \doi{https://doi.org/10.1007/s00158-008-0265-0}

\bibitem[{Zhou et~al.(2016)Zhou, Cao, and Lin}]{Zhou2016p1599}
Zhou S, Cao J, Lin J (2016) Theoretical analysis and experimental verification
  for improving energy harvesting performance of nonlinear monostable energy
  harvesters. Nonlinear Dynamics 86:1599--1611,
  \doi{https://doi.org/10.1007/s11071-016-2979-7}

\end{thebibliography}

\end{document}